\documentclass[11pt,preprint,flushrt]{aastex}

\usepackage{natbib,graphicx,amsmath,subfigure,color}
\def\eqq#1{Equation~(\ref{#1})}

\newcommand\ie{{\it i.e.\ }}

\newcommand{\veca}{\mbox{\boldmath $a$}}
\newcommand{\vece}{\mbox{\boldmath $e$}}
\newcommand{\vecE}{\mbox{\boldmath $E$}}
\newcommand{\vecg}{\mbox{\boldmath $g$}}
\newcommand{\vecG}{\mbox{\boldmath $G$}}
\newcommand{\vecM}{\mbox{\boldmath $M$}}
\newcommand{\vecQ}{\mbox{\boldmath $Q$}}

\newcommand{\vect}{\mbox{\boldmath $\theta$}}

\newcommand{\matR}{\mbox{$\bf R$}}
\newcommand{\matC}{\mbox{$\bf C$}}
\newcommand{\vecD}{\mbox{\boldmath $D$}}
\newcommand{\matD}{\mbox{$\bf D$}}
\newcommand{\matA}{\mbox{$\bf A$}}
\newcommand{\matI}{\mbox{$\bf I$}}
\newcommand{\bnab}{\boldsymbol{\nabla}}
\newcommand{\bnabg}{\boldsymbol{\nabla_g}}

\newcommand{\vecx}{\mbox{\boldmath $x$}}
\newcommand{\veck}{\mbox{\boldmath $k$}}
\newcommand{\vecv}{\mbox{\boldmath $v$}}
\newcommand{\vecb}{\mbox{\boldmath $b$}}

\newcommand{\likeli}{\mbox{$\mathcal{L}$}}
\newcommand{\lensfit}{{\sc lensfit}}

\begin{document}

\slugcomment{\$Revision: 2.6 $ $ \$Date: 2013/12/02 15:32:54 $ $,
  accepted to MNRAS}

\title{Bayesian Lensing Shear Measurement}

\author{Gary M. Bernstein}
\email{garyb@physics.upenn.edu}
\and
\author{Robert Armstrong}
\email{rearmstr@gmail.com}
\affil{Department of Physics \& Astronomy, University of Pennsylvania, 
209 S.\ 33rd St., Philadelphia, PA 19104}

\begin{abstract}
We derive an estimator of weak gravitational lensing shear from
background galaxy images that avoids noise-induced biases through a
rigorous Bayesian treatment of the measurement.  
The derived shear estimator disposes with the assignment
  of ellipticities to individual galaxies that is typical of previous
  approaches to galaxy lensing.
Shear estimates from
the mean of the Bayesian posterior are unbiased in the limit of large
number of background galaxies, regardless of the noise level on
individual galaxies.
The Bayesian
formalism requires a prior describing the (noiseless) distribution of the target
galaxy population over some parameter space; this prior can be
constructed from low-noise images of a subsample of the
target population, attainable from long integrations of a fraction of
the survey field.  We find two ways to combine this exact
treatment of noise with rigorous treatment of the effects of the
instrumental point-spread function and sampling.
The Bayesian model fitting (BMF) method assigns a likelihood of the
pixel data to galaxy models (e.g. Sersic ellipses), and requires the
unlensed distribution of galaxies over the model parameters as a prior.
The Bayesian Fourier domain (BFD) method compresses the
  pixel data
to a small set of weighted moments calculated after PSF correction in
Fourier space.  It requires the unlensed 
distribution of galaxy moments as a prior, plus derivatives of this
prior under applied shear.
A numerical test using a simplified model of a biased galaxy
measurement process demonstrates that the Bayesian formalism recovers
applied shears to $<1$ part in $10^3$ accuracy as well as providing
accurate uncertainty estimates.
BFD is the first shear
measurement algorithm that is 
model-free and requires no approximations or {\it ad hoc} assumptions
in correcting for the effects of PSF, noise, or sampling on the galaxy
images.  These algorithms are good candidates for attaining the
part-per-thousand shear inference required for hemisphere-scale weak
gravitational lensing surveys.
BMF has the
drawback that shear biases will occur since galaxies do not fit any
finite-parameter model, but has the advantage of being robust to
missing data or non-stationary noise.  
Both BMF and BFD methods are readily extended to use data from multiple
exposures and to inference of lensing magnification.  
\end{abstract}

\section{Introduction}
Gravitational lensing reveals the mass distribution of the Universe by
detecting deflections of photons in the gravitational potential
generated by the mass.  Along most lines of sight, we can best measure
the {\em gradient} of the deflection, characterized by an apparent shear \vecg\
and magnification
of background sources.\footnote{There are several  
  possible parameterizations for the gradient matrix in terms of
  shear.  We will leave this unspecified to emphasize that our method
  is valid for any choice of parameterization.}
The lensing shear is
detectable as a coherent alignment induced on nominally
randomly-oriented resolved background galaxies.  Reliable measurement
of this shear opens the door to a wealth of astrophysical and
cosmological information, including the most direct measures of the
dark components of the Universe.  See \citet{jainhoekstra} and
\citet{WeinbergDE} for recent reviews of the power of this weak
gravitational lensing technique.

The full power of the weak lensing technique can only be realized,
however, if we are able to infer the shear from real image data
without significant systematic error.  This apparently straightforward
measurement is complicated by several factors:
\begin{itemize}
\item The shear is weak, amounting to $\approx 2\%$ change in a
  galaxy's axis ratio on a typical cosmological line of sight.  In a full-sky
  experiment, it is possible to measure shear with
  statistical errors below 1 part in $10^3$ of this 2\%---
  systematic errors must be extremely small else they will dominate
  the error budget.
\item The galaxy is viewed through an instrument (possibly including
  the atmosphere) which convolves the lensed appearance
  with a point spread function (PSF) that typically
  induces larger coherent shape changes than the gravitational
  lensing, and can vary with time and with position on the sky.  This
  instrumental effect must be known and removed.
\item The received image of the galaxy is pixelized, meaning it has
  finite sampling.  Even if the sampling meets the Nyquist criterion
  so that the image is unambiguous, our shear extraction algorithm
  must handle the sampling and any other signatures of the detector.
\item The unlensed appearance of any individual galaxy is unknown, and
  galaxies have an infinite variety of intrinsic shapes.
  No finite parameterization can fully describe the unlensed galaxies.
\item The received image includes photon shot noise and additional
  noise from the 
  detector.  The shear inference must maintain exceptionally low bias
  even when targeting galaxies with signal-to-noise ratio $ S/N<10$--15 if
  we are to extract the bulk of the shear information available from
  typical optical sky images.
\end{itemize}
\citet{step}, \citet{step2}, \citet{great08}, and \citet{great10} document a series of
challenges in which the community was invited to infer the shear from 
simulated sky images, as a
means of assessing our abilities to measure shear in the face of the
above difficulties.  These publications also summarize the impressive
variety of 
techniques that have been proposed for shear inference.
A useful parameterization for errors in shear inference
is (using a simplified scalar notation) that the measured shear $\hat
g$ is related to the true shear via
\begin{equation}
\hat g = (1 + m) g + c.
\end{equation}
\citet{htbj} calculate that ambitious weak-shear surveys must obtain
multiplicative errors $|m|<10^{-3}$ to retain their full statistical
power.  The additive bias $c$, which can arise when the PSF or other
element of the analysis chain is not symmetric under 90\arcdeg\
rotation, must be kept below $\approx10^{-3.5}$, which is
100--300$\times$ smaller than the typical PSF ellipticity.  The
literature contains no demonstrations of robust shear algorithms
yielding $|m|<0.01$ at $S/N\approx 10$.  

\citet{fdnt} demonstrates the ability to attain $|m|<10^{-3}$ at high
$S/N$, overcoming all but the last of the problems itemized above.
This ``FDNT'' method has a rigorous formulation for noiseless data.  
It has, however, proven more difficult to derive a shear inference
that is rigorously correct in the presence of noise.
\citet{refregierbias} derive the lowest-order noise-induced bias in
galaxy-shape estimates that are produced via maximum-likelihood
fitting to parametric galaxy models.
But it is not clear that this correction yields
the necessary accuracy on real data.  
The most common
approach to biases induced by noise (or other systematic errors) has
been to use simulated sky data to infer $m$ and then apply a
correction to the real-sky result.  The accuracy of this approach is
of course limited by the extent to which the simulated data reproduces
the salient characteristics of the real sky. 
\citet{tomek} propose a somewhat more general scheme of calibrating
shape-measurement biases vs a few parameters of the galaxy and
measurement, {\it e.g.\/} the $S/N$ level, resolution, Sersic index, etc., and then
applying bias corrections on a galaxy-by-galaxy basis.  But
\citet{im3shape} note that this galaxy-by-galaxy bias correction is inaccurate
at low $S/N$, and it is better to determine a single overall
correction to the shear using
prior high-$S/N$ information on a random subsample of the source population.  

All these avenues lead us back to the situation of requiring prior empirical
high-$S/N$ information on the underlying source galaxy population in order to
produce the noise-bias corrections.  Once we need an empirical prior
on galaxy information, we should look to Bayesian techniques to
produce a rigorously correct shear estimator.  The \lensfit\ method of
\citet[\lensfit]{lensfit}
produces a Bayesian posterior distribution $P(\vece_i | \vecD_i)$ for
the ellipticity of galaxy $i$ given its pixel data $\vecD_i$,
assuming that galaxies take a known functional form and given a
prior distribution 
of \vece\ and the other parameters of the presumed galaxy model.
But in \lensfit\ the inference
of applied {\em shear} \vecg\ from an ensemble of these posterior densities
for galaxy {\em shapes} is done using some {\it ad hoc} weighting and
averaging schemes.  Testing of the \lensfit\ codes on simulated data
in \citet{lensfitcfh} yields $|m|\sim 0.1$ at
$S/N=10$, necessitating an empirical multiplicative correction to
shears derived from real data.

In this paper we will derive a rigorous Bayesian treatment of the
inference of weak {\em shear}, not just galaxy shapes, from pixel
data.  The general approach is outlined in Section 2.  Then we examine
three ways to apply this approach to real data: Section~\ref{method1}
examines Bayesian model-fitting (BMF) approaches and extends the \lensfit\ Bayesian
method to shear estimation.  Section~\ref{method2} then shows how to
apply Bayesian techniques without assuming a parametric model for
unlensed galaxies, yielding an adaptation of the
venerable \citet[KSB]{ksb} weighted-moment method that treats
convolution, sampling, and noise rigorously.  We consider this
Bayesian Fourier-domain (BFD) method to be our most promising
algorithm for high-accuracy shear inference.
In
Section~\ref{nulltests} we explore whether the FDNT method that is
successful at high $S/N$ can be embedded in a Bayesian framework for
accuracy at finite $S/N$.  In Sections~\ref{compare} and
\ref{conclude} we compare to some extant shear-measurement algorithms
and conclude.

\section{Bayesian shear estimate}
\subsection{General formulation}
\label{weakapprox}
We begin by assuming that we wish to infer the posterior distribution
of a constant shear \vecg\ from an image containing data \vecD\ that
can be subdivided into statistically independent subsets $\vecD_i$,
each covering a single galaxy $i\in\{1,\ldots,N_g\}$. For a prior probability $P(\vecg)$ of
the shear, the standard Bayesian formulation for the posterior is
\begin{equation}
\label{bayes1}
P(\vecg | \vecD) = \frac{P(\vecg) P(\vecD | \vecg)}{P(\vecD)}
 = P(\vecg) \prod_i \frac{P(\vecD_i | \vecg)}{P(\vecD_i)}.
\end{equation}
Our fundamental assumption is that the posterior $\ln P(\vecg |
\vecD)$ will be well approximated
by a quadratic Taylor expansion in \vecg, \ie\ the posterior will be
Gaussian in \vecg.  This assumption will fail when the number of
galaxies is small and the shear is poorly constrained, but should
become valid when we combine information from a large
number of galaxies on a weak shear.  In
Appendix~\ref{cubic} we give a prescription for including 3rd-order
terms in \vecg\ as perturbations,
since terms $O(g^3)$ are necessary if the estimate of \vecg\ is to be
accurate to $<1$ part in $10^{-3}$ for $g\approx0.03$.

The Bayesian formulation (\ref{bayes1}) produces an exact posterior
  distribution of the shear.  The full posterior distribution could be
  propagated into cosmological inferences, but most current analyses
  require instead an estimator $\hat g$ for the shear (and an uncertainty).
 By adopting a quadratic Taylor expansion for $\ln
  P(\vecg | \vecD)$ we implicitly assume that the maximum of the posterior distribution 
  is coincident with the mean of the posterior, so it would be natural
  to adopt the location $\bar{\vecg}$ of this posterior peak as a
  shear estimator.  Below we will consider whether this estimate is biased.

For quadratic order, we need the 6 scalars that
make up the scalar $P_i$, vector $\vecQ_i$, and symmetric $2\times2$
matrix $\matR_i$ defined as:
\begin{align}
P_i  & =  P(\vecD_i | \vecg=0) \notag\\
\label{pqr}
\vecQ_i & = \left. \bnabg P(\vecD_i | \vecg)\right|_{\vecg=0} \\
\matR_i & = \left. \bnabg \bnabg P(\vecD_i |
  \vecg)\right|_{\vecg=0} \notag \\
\Rightarrow\quad P(\vecD_i | \vecg) & \approx  P_i + \vecg \cdot
\vecQ_i + \frac{1}{2} \vecg \cdot \matR_i \cdot \vecg.
\label{pquad}
\end{align}
The dependence of the posterior on \vecg\ can now be expressed as
\begin{equation}
\label{Pg}
-\ln P(\vecg | \vecD) \approx {(\rm const)} - \ln P(\vecg) - \vecg \cdot \sum_i
\frac{\vecQ_i}{P_i}
+ \frac{1}{2} \vecg \cdot \left[ \sum_i \left(\frac{\vecQ_i \vecQ^T_i}{P_i^2}
-\frac{\matR_i}{P_i}\right) \right] \cdot
\vecg.
\end{equation}
If we presume that the data are much more informative than the prior
on $\vecg$, we may drop the $\ln P(\vecg)$ and find that the
posterior for the shear is Gaussian with covariance
matrix $\matC_g$ and mean $\bar{\vecg}$ defined via
\begin{align}
\label{Cg}
\matC_g^{-1} & = 
\sum_i \left(\frac{\vecQ_i \vecQ^T_i}{P_i^2} - \frac{\matR_i}{P_i}\right) \\
\label{barg}
\bar{\vecg} & =  \matC_g \sum_i \frac{\vecQ_i}{P_i}.
\end{align}
Note that we have {\em not} made any assumption about the Gaussianity
of the likelihood $P(\vecD_i | \vecg)$ for each galaxy, nor about any
priors, etc.;
only about the posterior of the applied shear \vecg.  

\subsection{Non-constant shear}
Before describing methods of calculating $P(\vecD_i |
\vecg)$ and its derivatives, we discuss generalization of
(\ref{Pg}) to non-constant shear fields.  Once the quantities $P_i,
\vecQ_i,$ and $\matR_i$ are calculated for each galaxy, the posterior
for any shear model can be calculated as long as the model predicts
shears in the regime where (\ref{pquad}) holds.  Consider a model with
parameter vector \veca\ which predicts shear values $\vecg_i(\veca)$ at
each galaxy.  
We can slightly modify our formulation and proceed as 
before to derive the posterior $P(\veca | \vecD)$ 
\begin{equation}
-\ln P(\veca | \vecD) \approx ({\rm const}) -\ln P(\veca) -
\sum_i \vecg_i(\veca)\cdot \frac{\vecQ_i}{P_i} + 
\frac{1}{2} \sum_i \vecg_i(\veca)\cdot  \left[ \sum_i \left(\frac{\vecQ_i \vecQ^T_i}{P_i^2} 
-\frac{\matR_i}{P_i}\right) \right] \cdot  \vecg_i(\veca),
\end{equation}
If $\vecg$ is a linear function of $\veca$ (and if the
prior is approximated as quadratic), then the
solution for the maximum-posterior $\veca$ is a closed-form matrix equation.
One potentially interesting application is
a Fourier decomposition $\vecg_i=\sum_\mu {\rm Re}(\veca_\mu e^{i{\bf k}_\mu
  \cdot {\bf x_i}})$.  
If the source galaxies are uniformly distributed on the plane and we
can make the approximation that
\begin{equation}
\sum_i \left(\frac{\vecQ_i \vecQ^T_i}{P_i^2} -\frac{\matR_i}{P_i}
\right) e^{i{\bf k}\cdot {\bf x}_i}=0 \quad {\rm for}\; {\bf k}\ne 0,
\end{equation}
then dependence of the posterior on the Fourier
coefficients becomes
\begin{align}
\ln P(\veca |\vecD)  \approx & \sum_\mu {\rm Re}\left[ \veca_\mu
\sum_i \frac{\vecQ_i}{P_i} e^{i{\bf k}_\mu \cdot {\bf x}_i}\right] \notag\\
  & + \frac{1}{4}\sum_\mu \veca_\mu \cdot {\matC}^{-1}_g \cdot \veca_\mu^\star, 
\end{align}
with $\matC^{-1}_g$ as from (\ref{Cg}).
This posterior separates into a Gaussian over each 2-component
$\veca_\mu$ with identical covariance $2\matC^{-1}_g$ on each of the real and
imaginary parts of each Fourier coefficient and a simple $2\times 2$
matrix solution for the most probable $\veca_\mu$.
A similarly simple
posterior could be derived for any decomposition of the shear field
into orthogonal functions.

Bayesian estimation of $N$-point correlations of a shear field should also be
similarly straightforward, at least in the limit where galaxy shape
noise and measurement noise are dominant over the sample variance of
the shear field.  We leave this derivation for future work.

\subsection{Bias of the Bayesian posterior}
Is the mean $\bar\vecg$ from \eqq{barg} an unbiased estimator of the
true shear \vecg?  The Bayes formalism does not guarantee that the
mean of the posterior is an unbiased estimator.  It does assure,
however that the posterior $P(\vecg | \vecD)$ converges to the input
value if the posterior is narrow, {\it i.e.} the Bayesian posterior is
not {\em wrong.}  

Define $ \overline{\vecb} = \langle \vecQ_i/P_i \rangle$ as the expectation of
the summand in \eqq{barg} over the population of target galaxies, and 
define 
\begin{equation}
\overline{\matA} = \left\langle \matA_i \right\rangle =
\left\langle \frac{\vecQ_i \vecQ^T_i}{P_i^2} -\frac{\matR_i}{P_i} \right\rangle.
\end{equation}
For each galaxy we can define $\delta\matA_i = \matA_i -
\overline{\matA}$ and 
$\delta \vecb_i = \vecQ_i/P_i - \overline{\vecb}$.

When we ignore the prior $P(\vecg)$ as
weakly informative relative to the data from a large number $N$ of
galaxies, 
the expectation of the mean $\bar \vecg$
of the posterior $P(\vecg | \vecD)$ in \eqq{barg} becomes
\begin{align}
\langle \bar\vecg \rangle & = \left \langle \left( N\overline{\matA} +
    \sum \delta\matA_i \right)^{-1} \left( N\overline{\vecb} + \sum
    \delta\vecb_i\right) \right\rangle \\
& = \left \langle
  \left( N\overline{\matA}\right)^{-1}
 \left( \matI +
    \frac{1}{N} \sum \matA^{-1}\delta\matA_i \right)^{-1} \left( N\overline{\vecb} + \sum
    \delta\vecb_i\right) \right\rangle \\
 & = \overline{\matA}^{-1} \overline \vecb 
 + \frac{1}{N}\overline{\matA}^{-1}
    \left[ \left\langle \delta\matA
     \overline{\matA}^{-1} \delta\matA \right\rangle
   \overline{\matA}^{-1} \overline{\vecb}
 - \left\langle \delta\matA
     \overline{\matA}^{-1} \delta\vecb \right\rangle \right]
 + O(N^{-2}).
\end{align}
The last line arises from expanding $(\matI+{\bf M})^{-1}
  \approx I - {\bf M} + {\bf M}{\bf M} - \ldots,$ and using $\langle
  \delta \vecb \rangle = 0,$ $\langle \delta\matA \rangle=0.$

As $N\rightarrow\infty$, $\matC_g\rightarrow (N\overline{\matA})^{-1}$
and the posterior narrows to a delta function
at $\bar g_\infty = \overline{\matA}^{-1}\overline{\vecb}$.  To the extent that our
quadratic approximation is valid, Bayes theorem demands that this
equal the true input \vecg.  We find then that the bias for a finite
set of galaxy is, to leading order in $1/N$, 
\begin{equation}
\label{gbias}
\langle \bar\vecg \rangle - \vecg \approx \frac{1}{N}
\overline{\matA}^{-1}
    \left[ \left\langle \delta\matA
     \overline{\matA}^{-1} \delta\matA \right\rangle \vecg
 - \left\langle \delta\matA
     \overline{\matA}^{-1} \delta\vecb \right\rangle \right].
\end{equation}
The first bracketed term is roughly a multiplicative bias (not quite,
because the distributions of \matA\ will depend weakly on \vecg).  If
we have $\overline{\matA}^{-1} \delta \matA = O(1)$, then this
multiplicative bias on shear is $O(1/N)$.  This bias will be
$\sim\sqrt{N}$ smaller than the statistical error $\sigma_g/\sqrt{N} \approx
0.3/\sqrt{N}$ that arises from shape noise in a weak lensing measurement.

The forms of \vecb\ and \matA\ suggest that $(\delta \vecb)^2 \sim
\overline{\matA} \sim \sigma_g^{-2}$ in magnitude, so the second term can be expected to scale as
$\sigma_g / N$.  In addition, this term involving covariance between \matA\ and
\vecb\ will vanish by symmetry at $\vecg=0$ if the PSF and noise are
isotropic.  Therefore we expect this bias to have another factor of
$g$ or of $e_\star$, the PSF ellipticity, in front of it, leading to
an additive bias of perhaps $e_\star \sigma_g / N$.  This will again
always be below shape noise and insignificant for shear statistics
constrained by $N\gg 10^4$ galaxy measurements.

These terms do represent a kind of noise bias on the mean of the shear
posterior taken as a shear estimator.  The principal difference from
previous techniques, however, is that the size of this bias
scales as the square of the measurement error on the {\em shear}, and
is not affected by measurement errors on individual galaxy shapes.

\subsection{Galaxy descriptors}
Implementation of the Bayesian method depends on assigning a probability
$P(\vecD_i | \vecg)$ to the pixel data given the shear.  To do so  we
introduce some finite set $\vecG_i$ of quantities describing the
appearance of galaxy $i$.  We must be able to assign a likelihood
$\likeli(\vecD | \vecG)$ of the pixel data \vecD\ being produced by a
galaxy with properties \vecG.  We also need to know the distribution
of $\vecG$ for the true galaxy population viewed through shear \vecg. Then we can assign
\begin{equation}
P(\vecD_i | \vecg) = \int d\vecG\, \likeli(\vecD_i | \vecG) P(\vecG |
\vecg).
\end{equation}
The derivatives with respect to \vecg\ needed to define $\vecQ_i,$ and $\matR_i$ in
(\ref{pqr}) all propagate purely to the prior $P(\vecG | \vecg)$
inside the integral.

In the next two sections, we will consider first a model-fitting
approach, in which the galaxy properties \vecG\ are assumed to predict
all of the observed pixel data values; and a model-free scheme, in
which the pixel data are compressed to a smaller set \vecM\ of
moments that will serve as our galaxy properties.  In either case, the
essential requirements are that:
\begin{enumerate}
\item We have a rigorous means to assign a likelihood $\likeli(\vecD |
  \vecG)$, and
\item The distribution of real galaxies' properties
  \vecG\ changes under application of shear \vecg, and that we can
  determine this dependence $P(\vecD|\vecg)$---{\it i.e.} there
  is a detectable and known signature of shear upon the galaxy population.
\end{enumerate}

\section{$P(\vecD_i | \vecg)$ via model fitting}
\label{method1}
\subsection{General Formulation}
If galaxy $i$ is assumed to be fully described by a model with a finite
number of parameters \vecG, all the instrumental signatures are known,
and there is a known noise model for the pixel data, then one can 
calculate the likelihood of the full pixel data vector $\likeli(\vecD_i |
\vece_i^o, \vecx_i, \vect_i)$.  Here we divide \vecG\ into three subsets:
\begin{itemize}
\item $\vece_i^o$ is a vector of observed parameters that are altered
  under the action of lensing shear \vecg, {\it i.e.\/} the ellipticity
  of the galaxy model.  We presume that there is an exactly known
  transformation under the action of shear from intrinsic source
  parameters to the observed parameters, $\vece^o = \vece^s \oplus
  \vecg$. Such is the case, for example, when $\vece$ is the
  2-component ellipticity of a galaxy with self-similar elliptical
  isophotes and \vecg\ is any of the common representations of the
  shear linear transformation matrix. We leave the form of this transformation free at this point
  to accommodate any convention for the
  definition of the ellipticity $\vece$ and the shear $\vecg$.  We do,
  however, require that the transformation be reversible:
  $(\vece \oplus \vecg) \oplus (-\vecg) = \vece$.\footnote{\vece\
    need not even be 2-dimensional: galaxy models with multiple
    components of different ellipticities might, for example, have 4
    or more elements of \vece.  The key is that their transformation
    under shear must be known.}
\item $\vecx_i$ is the center of the galaxy, or more generally any
  parameters whose prior distribution $P(\vecx_i)$ can be taken as uniform both
  before and after the application of shear to the sky.  We hence will
  not make $\vecx_i$ an argument of our priors.
\item $\vect_i$ are other parameters of the model, which we take to be
  invariant under the action of shear on the image.  Examples would be
  the half-light radius, surface brightness, and Sersic index of a
  simple elliptical Sersic-profile galaxy model, {\it e.g.} as used in
  \citet{lensfit}.  
\end{itemize}
The shear-conditioned probability for single galaxy becomes
\begin{equation}
P(\vecD_i | \vecg) = \int d\vece^o_i\, d\vecx_i\,d\vect_i \likeli(\vecD_i
  |\vece^o_i, \vecx_i, \vect_i) P(\vece^o_i, \vect_i | \vecg),
\end{equation}
where $P(\vece^o_i, \vect_i | \vecg)$ is the prior distribution the
galaxy parameters given the shear.  The shear enters the posterior
only through this term.  Conservation of probability under shear
requires that
\begin{equation}
\label{prior}
P(\vece^o, \vect | \vecg) = P_0(\vece^o \oplus -\vecg, \vect)
\left| \frac{d\vece^s}{d\vece^o} \right|_{-\vecg},
\end{equation}
where $P_0(\vece, \vect)$ is the {\em unlensed} distribution of galaxy
ellipticities, and the last term is the Jacobian of the ellipticity
transformation $\vece^s = \vece^o \oplus -\vecg$.  For an isotropic
Universe, the unlensed prior should be a function only of the
amplitude $e=|\vece|$, not the orientation.  Given the unlensed prior,
we can make a Taylor expansion in the shear:
\begin{equation}
\label{priorTaylor}
 P_0(\vece^o \oplus -\vecg, \vect)
\left| \frac{d\vece^s}{d\vece^o} \right|_{-\vecg} =
P_0(e^o, \vect)+ \vecg \cdot \vecQ(\vece^o, \vect) + \frac{1}{2} \vecg
\cdot \matR(\vece^o, \vect) \cdot \vecg + O(g^3).
\end{equation}
Once the transformation $\vece\oplus\vecg$ is specified, the functions
\vecQ\ and \matR\ can be derived in terms of $P_0(e,\vect)$ and its
first two derivatives with respect to $e$.
The quantities needed from each galaxy for the Bayesian
posterior in \eqq{Pg} are
\begin{equation}
\label{int1}
\left\{ \begin{array}{c}
P_i \\ \vecQ_i \\ \matR_i
\end{array} \right\} 
\equiv  \int d\vece^o_i \, d\vecx_i\,d\vect_i \likeli(\vecD_i
  |\vece^o_i, \vecx_i, \vect_i) 
\left\{\begin{array}{c}
P_0( e^o_i, \vect_i) \\ \vecQ( \vece^o_i, \vect_i) \\ \matR(
\vece^o_i, \vect_i)
\end{array} \right\}.
\end{equation}
Note that $P_i$ would be the Bayesian evidence for $\vecD_i$ in the absence of
lensing shear.

The operative procedure for obtaining the Bayesian shear estimate is:
\begin{enumerate}
\item Determine the unlensed prior $P_0(e,\vect)$ from a high-$S/N$
  imaging sample, and perform the Taylor expansion in
  \eqq{priorTaylor} knowing the ellipticity transformation equation.
\item For each observed galaxy, compute the six distinct integrals in
  (\ref{int1}) with the likelihood over the ellipticity 
  and structural parameters to get $P_i, \vecQ_i,$ and $\matR_i$.
\item Sum over galaxies to obtain $\matC_g^{-1}$ in \eqq{Cg}.  
\item Sum over galaxies to obtain the shear estimate $\bar{\vecg}$ as
  in \eqq{barg}.  The shear posterior has this mean (and maximum) and
  covariance matrix $\matC_g$.
\end{enumerate}
Note that the division by $\matC_g^{-1}$ occurs after the summation
over galaxies, \ie\ we do not generate ellipticity or shear estimators
on a galaxy-by-galaxy basis.  
Also note that for galaxies with very noisy
data, $\likeli(\vecD_i)$ is very weakly dependent on galaxy
properties and hence on shear: $\bnabg P(\vecD_i | \vecg)\rightarrow 0.$
Equations~\ref{pqr} then show that
\vecQ\ and \matR\ both tend to zero.  The
low-$S/N$ galaxy hence has
no influence on the shear likelihood.  It will therefore be unnecessary
to make cuts on galaxy size or $S/N$ ratio to obtain a successful
measurement.  As long as the source selection is made on a quantity
(such as total flux) that is
unaffected by galaxy shape or lensing shear, we avoid selection biases.

The only approximation made in this derivation was that of weak shear,
namely that \eqq{priorTaylor} is accurate over the range of \vecg\
permitted by the data.  If the unlensed ellipticity distribution is
characterized by a scale of variation $\sigma_e$, this means we are assuming that $g\ll \sigma_e$.  

Extension to multiple exposures of the same galaxy is trivial: the
pixel data $\vecD_i$ for galaxy $i$ may be the union of $N$ distinct
exposures' information, $\vecD_i =
\{\vecD_{i1},\vecD_{i2},\ldots,\vecD_{iN}\}.$  Assuming that different
exposures have statistically independent errors, we have
\begin{equation}
\likeli(\vecD_i | \vece_i^o, \vecx_i, \vect_i)= \prod_{j=1}^N
\likeli(\vecD_{ij} | \vece_i^o, \vecx_i, \vect_i).
\end{equation}
If the exposures are in different filter bands, then the most general
formulation is that $\vece_i$ and $\vect_i$ are the unions of
distinct structural parameters $\vece_{ij}$ and $\vect_{ij}$ for each
band $j$, and the prior must
specify the joint distribution of galaxies' appearances in all observed bands.

An alternative formulation to the above is to integrate over the
source ellipticity $\vece^s$ instead of the lensed ellipticity
$\vece^o$, which puts the derivatives with respect to
shear in the likelihood instead of the prior:
\begin{equation}
P(\vecD_i | g) = \int d\vece^s_i\, d\vecx_i\,d\vect_i \likeli(\vecD_i
  |\vece^s_i\oplus\vecg, \vecx_i, \vect_i) P_0(e^s_i, \vect_i | \vecg).
\end{equation}
In this case the quantities needed for the
shear posterior are
\begin{align}
P_i & \equiv  \int d\vece^s_i \, d\vecx_i\,d\vect_i \likeli(\vecD_i
  |\vece^s_i, \vect_i) P_0( e^s_i, \vect_i) \notag \\
\label{int1b}
\vecQ_i & \equiv  \int d\vece^s_i \, d\vecx_i\,d\vect_i 
\left[ \left.\bnabg \likeli(\vecD_i  |\vece^s_i\oplus g, \vect_i) \right|_{\vecg=0}\right] P_0( e^s_i, \vect_i) \\
\matR_i & \equiv  \int d\vece^s_i \, d\vecx_i\,d\vect_i 
\left[ \left.\bnabg \bnabg \likeli(\vecD_i  |\vece^s_i\oplus g,
  \vect_i) \right|_{\vecg=0}\right] P_0( e^s_i, \vect_i) \notag
\end{align}
We would expect
Equations~(\ref{int1}) to be the more computationally
efficient approach, since the derivatives of the prior with respect to
shear can be pre-calculated once, whereas (\ref{int1b})
require calculating 5 shear derivatives of the likelihood for every
target galaxy.  Furthermore the isotropy and parity symmetries of the
unlensed sky simplify the shear derivatives of the prior.

\subsection{Simplified demonstration}
\subsubsection{Adopted model}
A numerical test with a simplified model demonstrates the accuracy and feasibility of the Bayesian
shear estimates in the presence of highly non-Gaussian likelihoods for
the ellipticities of individual galaxies.   We take a minimal set of
galaxy properties to be $\vecG = \vece^o$, the two components of a
(post-lensing) true galaxy shape.  We take an absolute minimal data vector
$\vecD_i=\vece^m$ for each galaxy to be a measurement of the
ellipticity, ignoring centroid and any structural
parameters. For each galaxy we:
\begin{itemize}
\item Draw a source ellipticity $\vece^s$ from an isotropic
  unlensed distribution
  truncated to $e^s = |\vece^s|<1$ and defined by
\begin{equation}
\label{bobprior}
P_0(e^s)\propto \left[1-(e^s)^2\right]^2 \exp\left[-(e^s)^2/2\sigma_{\rm prior}^2\right].  
\end{equation}
The additional factor of $(1-(e^s)^2)^2$ atop the Gaussian ensures the prior has two
continuous derivatives at the $|e|<1$ boundary. 
\item Generate a lensed ellipticity $\vece^o = \vece^s \oplus
  \vecg$ for a constant shear \vecg, using the full non-Euclidean
  transformation for ellipticity under shear {\it e.g.\/} as described
  by \citet{SS97}.
\item Obtain a measurement $\vece^m$ by drawing from a
    Gaussian distribution with a variance of $\sigma^2_m$ per axis.
    This measurement error is made non-Gaussian by truncation to $|
    \vece^m | < 1$.   Furthermore we model biases in the measurement
    process by centering the Gaussian at a value $\vece_{\rm ctr} =
    (1+m_{\rm bias})\vece^o + \vece_{\rm bias}$ for some
    multiplicative error $m_{\rm bias}$
    and additive error $\vece_{\rm bias}$.  In the numerical tests
    below, we adopt $m_{\rm bias}=-0.1$ and $\vece_{\rm bias}=(0.03,0)$.
\end{itemize}
 The likelihood $\likeli(\vecD_i |
\vecG) = \likeli(\vece^m_i | \vece^o)$ is a known 
truncated, biased function. The measurement
error distribution is asymmetric with non-zero mean, and strongly dependent
upon the intrinsic ellipticity---characteristics which induce biases in
most extant shear-estimation methods. 

\subsubsection{Integration algorithm}
The integrals in (\ref{int1}) are evaluated using a grid-based 
approach adapted from \citet{lensfitcfh}. 
We define a set of sampled points, starting by
evaluating the integrand of $P_i$ in (\ref{int1}) at a random point in
$\vece^o$.  We construct a square grid with
initial resolution $\Delta e$ centered on the initial sample.  We then iterate this process:
\begin{enumerate}
\item Evaluate the integrand at all grid points that neighbor existing
  members of the sample.
\item Determine the maximum integrand value $I_{\rm max}$ among all
  sampled points.
\item Discard any sampled point with integrand $< t_{\rm min} I_{\rm max}$.
\end{enumerate}
We iterate this process until no new samples are added on an iteration.

If the number of surviving above-threshold samples is $\ge N_{\rm
  min}$, we halt the process.  Otherwise we decrease the grid spacing
$\Delta e$ by
a factor of $\sqrt{2}$ by adding a new grid point at the center of
each previous grid square. We then repeat the above iteration.  The grid is refined until we 
reach the desired number of above-threshold samples or until we reach a minimum resolution.  
We can now calculate the integrals in (\ref{int1}) for each galaxy by summing 
over the sampled points $\vece_{ij}$:
\begin{equation}
P_i = \sum_j (\Delta e)^2 P(\vece^m_i |\vece^o_{ij}) P_0( e^o_{ij}).
\end{equation}
and evaluate $\vecQ_i$ and $\matR_i$ by summing over the same 
samples, reusing the calculated $\likeli(\vece^m|\vece^o)$,
changing only the last terms as per the integrands in (\ref{int1}).

\subsubsection{Results}
For all tests we fix $\sigma_{\rm prior}=0.3$, $\vecg=(0.01,0)$, and
$N_{\rm min}=50$.
For each parameter set we simulate at least $0.75\times 10^9$ galaxies to reduce 
the shape noise below our desired accuracy.  

\begin{figure}[ht]
\centering
\plottwo{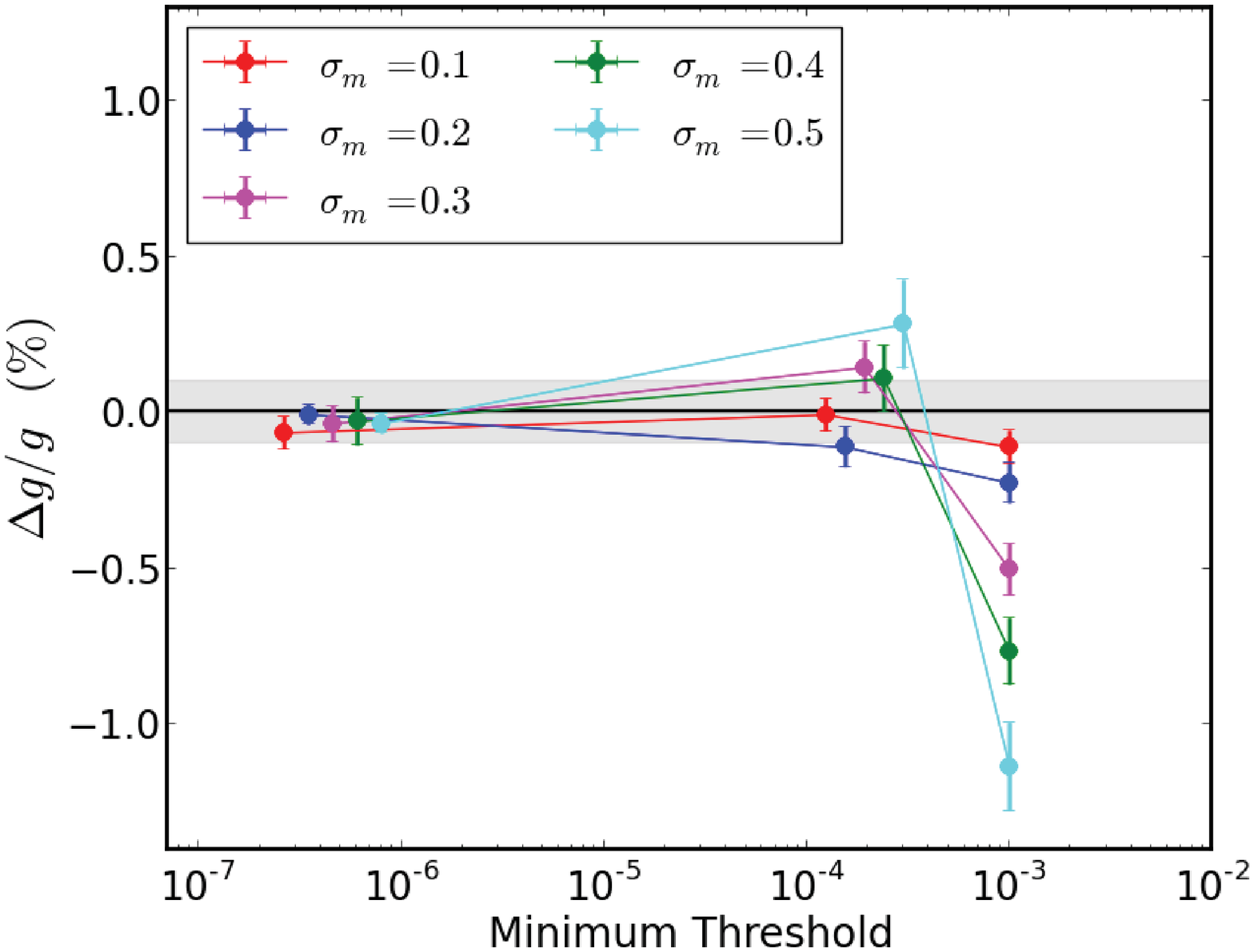}{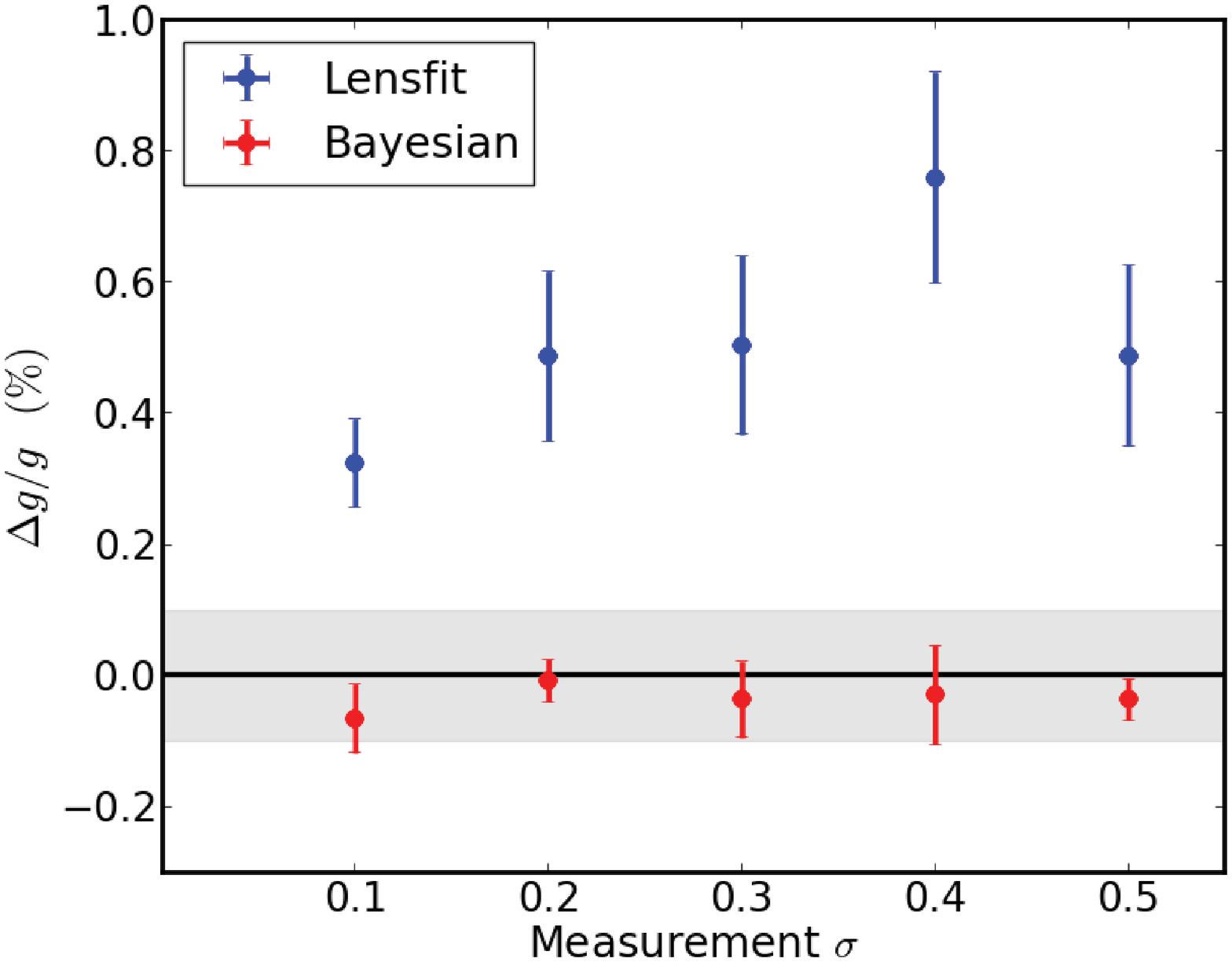}
\caption{
At left, the accuracy to which we are able to recover 
the reduced shear $\vecg$ in our toy model is shown as a function 
of probability threshold $t_{\rm min}$ (below the maximum) above which we keep points
for the integration.  The different curves show the 
accuracy for different ellipticity measurement errors $\sigma_m$.  The shaded region 
shows the desired level of accuracy $m=\Delta g/g < 10^{-3}$ on $\vecg$. For $t_{\rm
  min}<10^{-4}$ we find shear estimation with the desired accuracy.
At right are the results for $t_{\rm min}=10^{-7}$ showing the desired
accuracy is obtained even for quite large, biased, non-Gaussian
measurement errors in the model.  Estimation of shear for the
simulated data from the
\lensfit\ formulae of \citet{lensfit} yields biases up to $\approx
10\times$ larger than our targets.
}
\label{recovery}
\end{figure}

The left plot in Fig.~\ref{recovery} shows the relative error $| \bar \vecg - \vecg |
/ |\vecg|$ for different values of $t_{\rm min}$ and $\sigma_m$.  
The shaded region is the desired relative shear error of
$|m|<1\times10^{-3}$.  
We recover $\vecg$ to this desired precision for all 
values of $\sigma_m$ as long as $t_{\rm min}$ is below $10^{-4}$, even
when the shear measurement error is as large as $\sigma_m=0.5$, as one
might obtain for real galaxies with $S/N\approx 4$.
The number of likelihood evaluations per galaxy is between 50 and 200
for all cases tested.

We also use a jacknife method to measure the statistical uncertainty
in the shear estimator (\ref{barg}) from each simulation. The measured
uncertainty in the shear is found to agree (to 1--2\%) with the
covariance matrix $\matC_g$ derived in (\ref{Cg}) from the Bayesian
framework.

The toy model illustrates the validity of the weak-shear Bayesian
formalism in the face of non-Euclidean shear transformations and messy
(but known) shape measurement errors.  

The right-hand plot in Figure~\ref{recovery} plots the
  Bayesian shear measurement error vs the per-galaxy ellipticity measurement
  error $\sigma_m$, for the case $t_{\rm min}=10^{-7}$.  For
  comparison we plot in blue the shear estimated for the same
  simulated data using the \lensfit\ estimator described in
  section~2.5 of \citet{lensfit}.  A galaxy weighting function is
  allowed in \lensfit: we assume equal weighting for our simulated
  galaxies. The fully Bayesian shear estimator attains the desired
  $\Delta g/g < 0.1\%$ while the \lensfit\ estimator does not. 
 Note
  that \citet{lensfitcfh} present a different \lensfit\ estimator
  which has been applied to the Canada-France-Hawaii Lens Survey.

\subsection{Errors from unrecognized structural parameters}
\label{hidden}
The primary difficulty of implementation of these Bayesian shear
measurement methods will be the need to construct high-dimensional
priors.  We will be tempted to reduce the dimensionality of the galaxy
parameter space.  Under what circumstances can we omit a parameter
from our Bayesian calculation and still obtain a rigorously correct
result?   Consider the simple case where a galaxy property set \vecG\
is supplemented by a parameter $\alpha$ which can
take discrete values $\alpha_1, \alpha_2, \ldots$ with probability
$p_1, p_2, \ldots.$  The posterior contribution from
a single galaxy is
\begin{equation}
\label{correct}
P(\vecD_i | \vecg) \propto 
\int d\vecG\,
\sum_j \likeli(\vecD_i|\vecG,\alpha_j)P(\vecG |\vecg,\alpha_j) p_j.
\end{equation}  
If we are unaware of this parameter or choose to ignore it, we will
have a prior marginalized over $\alpha$ and probably assign a
likelihood that is also an average over the $\alpha$ cases.  Our
posterior calculation will then yield
\begin{equation}   
\label{avg}
P(\vecD_i | \vecg) \propto \int d\vecG\,
\left(\sum_j p_j \likeli(\vecG, \alpha_j)\right) 
\left(\sum_j p_j P(\vecG |\vecg,\alpha_j)\right).
\end{equation}
This can differ from the correct (\ref{correct}) unless either the prior or data likelihood 
is independent of $\alpha$.  In other words: {\em the prior must specify
the distribution of all galaxy parameters that the likelihood depends
upon.}

\begin{figure}[ht]
\centering
\includegraphics[width=3.2in]{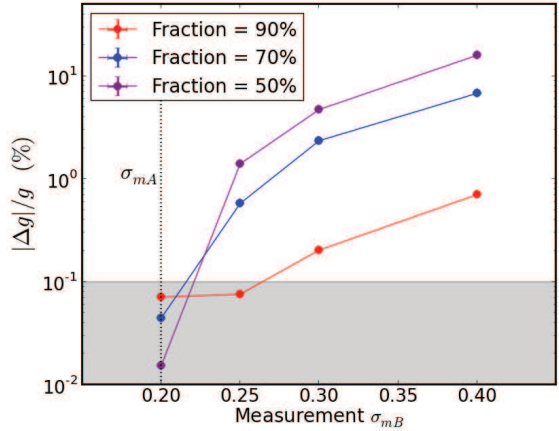} 
\caption{The multiplicative shear bias induced by the unrecognized
  presence of two galaxy populations A and B is plotted vs the
  measurement error $\sigma_{mB}$ of the B population.  The A population has
  measurement error $\sigma_{mA}=0.2$.  The intrinsic ellipticity
  dispersions for A and B galaxies are 0.1 and 0.3, respectively.  The
  shear bias from the unrecognized structural parameter vanishes when
  the error distributions match and $\likeli(\vecD | \vece^o)$ becomes
  independent of galaxy type.  It is also reduced as the fraction
  $\alpha_B$ of the population in Type B is decreased.}
\label{pop}
\end{figure}

We illustrate such bias by dividing the galaxies in our toy model into two 
populations $A$ and $B$ with distinct $\sigma_{\rm prior}$ and $\sigma_m$.  
Galaxies are assigned to Type B at random with some probability $\alpha_B.$
An observer ignorant of the existence of Type A
and Type B galaxies would infer a prior and a measurement-error
distribution that are found by averaging over the 
full population as in~(\ref{avg}).  Fig.~\ref{pop} shows the
accuracy of the Bayesian shear estimate under these (mistaken) assumptions.
We choose $\sigma_{{\rm prior}A}=0.1$, $\sigma_{{\rm prior} B}=0.3$,
$\sigma_{mA}=0.2$, and vary $\sigma_{mB}$.  
As expected a shear bias (up to 15\%) appears as
$\sigma_{mB}$ becomes more distinct from $\sigma_{mA}.$  The shear
bias decreases when Type B galaxies become rarer and when their
measurement error (hence likelihood function) become indistinguishable
from Type A.

For this Bayesian model-fitting method, an example of a galaxy parameter that
can be ignored is color, which does not affect the likelihood function
of single-band pixel data as long as the other parameters $\{\vece,
\vecx, \vect\}$ completely specify the single-band appearance of the
galaxy.  The accuracy of the Bayesian inference may be improved,
however, if we can include measured color in the data vector and
include color dependence in the prior, {\it e.g.\/} by distinguishing
late- and early-type galaxies.

In model-fitting, ignoring parameters that {\em do} affect the
likelihood of pixel values can produce bias.  An example would be
assuming a fixed bulge-to-disk ratio when fitting a population of
galaxies that has varying ratios.  

Unfortunately there is no known finite parameterization for real
galaxies' appearances, and hence it is not possible to create a
formally correct Bayesian model-fitting shear method for real
galaxies.  The means by which incomplete galaxy models can bias shear
measurements are illustrated and explained by \citet{vb},
\citet{MelchiorViola}, and \citet{fdnt}.

The potential for shear biases from unrecognized galaxy
characteristics is not limited to model-fitting methods.  
For example
in the model-free moment-based method described in the next
section, the Bayesian formalism will break down if the likelihood of the
moments depends on the detailed structure of the underlying galaxy.
As noted above, we need to marginalize over any galaxy parameter that
appears in the likelihood expression for the data.

\section{$P(\vecD_i | \vecg)$ via data compression}
\label{method2}
\subsection{General Formulation}
Acknowledging that we cannot construct a
model of galaxy structure with which to predict all pixel
values, we can instead compress the pixel data into a small number of
quantities that carry most of the information on any shear applied to
the source. We will call the compressed quantities ``moments'' since we propose that
they be intensity-weighted moments.  This method will hence resemble
the venerable \citet{ksb} shear-measurement methodology, but with
critical changes to eliminate the approximations inherent to KSB and a
rigorous Bayesian formulation to eliminate noise-induced biases.

We choose to reduce the pixel data for any galaxy image to a small vector
$\vecM=\{M_1,M_2,\ldots,M_m\}$ of derived quantities which we will
select to be sensitive to shear.
The \vecM\ must be chosen such that one can
propagate the pixel noise model to the moment vector.  If $\vecM_i$
are the moments calculated from the pixel data for galaxy $i$, we must
be able to assign a likelihood $\likeli(\vecD_i | \vecG) = \likeli(\vecM_i | \vecM)$ 
of producing the measured moments given that the galaxy has true (noiseless)
moments \vecM.  In this case the contribution to the shear posterior
from galaxy $i$ is 
\begin{equation}
\label{posterior2}
P(\vecD_i | \vecg) \propto \int d\vecM\, d\vecx\, \likeli(
\vecM_i | \vecM) P(\vecM | \vecx, \vecg)
= \int d\vecM\, \likeli(
\vecM_i | \vecM) \int d\vecx\, P(\vecM | \vecx, \vecg),
\end{equation}
where $P(\vecM | \vecx, \vecg)$ is the prior distribution of moments
of a galaxy centered at \vecx\ given a local shear \vecg.  The Taylor
expansion of the prior can be written as
\begin{equation} 
\label{priorTaylor2}
\int d\vecx\, P(\vecM | \vecx, \vecg) = P_0(\vecM) +
\vecg \cdot \vecQ(\vecM) + \frac{1}{2} \vecg
\cdot \matR(\vecM) \cdot \vecg + O(g^3).
\end{equation}
and then the quantities needed from each galaxy for the Bayesian
posterior in \eqq{Pg} are
\begin{equation}
\label{int1c}
\left\{ \begin{array}{c}
P_i \\ \vecQ_i \\ \matR_i
\end{array} \right\}
 \equiv  \int d\vecM \, \likeli(\vecM_i | \vecM)
\left\{ \begin{array}{c}
 P_0(\vecM) \\ \vecQ( \vecM) \\ \matR( \vecM)
\end{array} \right\}
\end{equation}

Now the operative procedure for obtaining the Bayesian shear is:
\begin{enumerate}
\item Determine the prior $P(\vecM | \vecx, \vecg)$ from a high-$S/N$
  imaging sample, and perform the Taylor expansion in
  \eqq{priorTaylor2}.  Here $\vecx$ is the vector from the coordinate
  origin of the moments to the center of the galaxy.
 The
  derivatives of this prior under shear must be obtained 
  by simulating the action of shear on each member of the high-$S/N$
  ensemble defining the prior.   
 Appendix~\ref{momentcalcs} shows that the derivatives of the moments
 \vecM\ are equal to a set of higher-order moments, which can hence be
 measured directly from the high-$S/N$ sample.
There is no longer any explicit
  ellipticity parameter describing each galaxy to represent the
  shear-dependent aspects of that galaxy.

\item For each observed galaxy, measure the moments $\vecM_i$ of that
  galaxy about some pre-selected coordinate origin and determine the likelihood
  function $\likeli(\vecM_i | \vecM)$.  We have no further need of the
  pixel data for the galaxy after this compression.
\item Compute the six distinct integrals in
  (\ref{int1c}) over the 
  moment space $\vecM$ to get $P_i, \vecQ_i,$ and $\matR_i$.  This is
  the computationally intensive step.
\item Sum over galaxies to obtain $\matC_g^{-1}$ in \eqq{Cg}.  
\item Sum over galaxies to obtain the shear estimate $\bar{\vecg}$ as
  in \eqq{barg}.  The shear posterior has this mean (and maximum) and
  covariance matrix $\matC_g$.
\end{enumerate}

\subsection{A specific choice of data compression}
To make things more concrete, we propose that the compressed
quantities \vecM\ be intensity-weighted moments of the galaxy image.
We will evaluate these in Fourier domain where, in the absence of
aliasing, the exact correction for the effect of the point spread function
(PSF) on the observed galaxy image is a simple division.  This yields
our Bayesian Fourier-domain (BFD) algorithm.
The moment vector could be
\begin{equation}
\label{moments}
\left\{ \begin{array}{c}
M_I \\
M_x \\
M_y \\
M_r \\
M_+ \\
M_\times
\end{array}
\right\} = \int d^2k\, \frac{\tilde I^o(\veck)}{\tilde T(\veck)}
W(|\veck^2|)
\left\{ \begin{array}{c}
1 \\
ik_x \\
ik_y \\
k_x^2 + k_y^2 \\
k_x^2 - k_y^2 \\
2 k_x k_y
\end{array}
\right\}.
\end{equation}
Here $\tilde I^o(\veck)$ and $\tilde T(\veck)$ are the Fourier
transforms of the observed image and the PSF, respectively, and $W(|\veck^2|)$
is a window function applied to the integral to bound the noise, in
particular confining the integral to the finite region of $k$ in which
$\tilde T(\veck)$  is non-zero, as detailed in \citet{fdnt}.  There is
a specific weight that will offer optimal $S/N$ for a given
galaxy$+$PSF pair, but any properly bounded weight with finite 2nd
derivatives produces a valid shear method.  One can select a single
weight function for a full survey and retain good $S/N$ as long as the
PSF does not vary widely in size.
In real data
the integrals revert to sums over the $k$-space values sampled by a
discrete Fourier transform (DFT) of the pixel data $D_i$.  

The first motivation for this choice is that these moments are
linear in the observed pixel data and therefore meet the requirement
that we be able to construct a moment noise model $\likeli(\vecM_i |
\vecM)$  from the known pixel noise model.  
In fact if the noise
in the pixels is independent of the pixel values, then the covariance
matrix $\matC_i$ of the moments is independent of the value of \vecM, and also
independent of any other properties of the galaxy.  Such is the case
for the background-limited conditions typical of faint-galaxy imaging.  If
the noise is stationary, than all galaxies with the same PSF will have
the same $\matC_i$.  
Even if the pixel noise is not Gaussian, the central limit
theorem implies that the moments, which are sums over many independent
pixels,
will tend toward a Gaussian distribution.  We therefore can take
\begin{equation}
\label{gausslikelihood}
-2 \ln \likeli(\vecM_i | \vecM) = \ln |2\pi \matC_i| +
(\vecM_i-\vecM)\cdot \matC_i^{-1} \cdot (\vecM_i - \vecM).
\end{equation}
Furthermore if the PSF is (nearly) isotropic,
then the azimuthal symmetries of our chosen basis set guarantee that
$\matC_i$ will be (nearly) diagonal, with the exception of a
covariance between the two monopole moments $M_I$ and $M_r$.  

One departure from common shear-measurement practices is that
calculating $\vecM_i$ from pixel data for $i$ does not involve any iterative
procedures such as centroiding, since iteration usually produce non-analytic likelihoods.  In
Section~\ref{nulltests} we look for feasible approximations to the
propagation of errors into some iterative compressed quantities.

We are also tempted to
reduce the dimensionality of our prior and speed up
marginalization by working with the normalized ratios $M_x/M_I$,
$M_y/M_I$, $M_+/M_r$, and $M_\times/M_r$ that figure prominently in KSB, then
dropping $M_I$ and $M_r$ from our data vector.  The probability distribution for such
ratios of Gaussian deviates is known \citep{MelchiorViola}, but depends
on the mean value and variance of the denominator,  so the moment
likelihood function would still depend on
$M_I$ and $M_r$.  As per the discussion in Section~\ref{hidden}, we
would still need to include these quantities in the
prior and marginalize over them, foiling our plan to simplify the
prior.  We choose the straightforward path of compressing the pixel
data to un-normalized moments.  

What is the motivation for choosing this particular set of linear
moments?  Any choice of well-defined compressed data
vector $\vecM$ will yield
a {\em valid} Bayesian shear estimator under this method, but the choices will
differ in the {\em precision} to which they determine the shear \vecg\ from
a given galaxy sample.  We want \vecM\ to include quantities that
unambiguously capture most of the impact of shear upon the pixel image.
The quadrupole moments $M_+$ and
$M_\times$ respond at first order to shear.  They are also sensitive
at first order to the galaxy size and flux, and at second order to
translation of the galaxy image, so we add the monopole and dipole
moments $M_I$, $M_r$, $M_x$, and $M_y$, sensitive to flux, size, and translation in
first order, to yield less degenerate shear information for each galaxy. The inclusion of $M_r$
furthermore would admit generalizing the Bayesian formalism to infer the
weak lensing magnification $\mu$ along with the 2 shear components \vecg.

\subsection{Generating the Prior}
We defer testing of a full implementation of this method for a later
publication, but we do outline the necessary steps and possible
economies.  

The biggest astrophysical challenge will be to produce the prior,
which is a function of the six \vecM\ components---which can be
reduced to 5 dimensions
because the isotropy of the Universe requires the unlensed prior to be
invariant under coordinate rotation.   At each point in the
5-dimensional space, we require 5 shear derivatives of the prior as
well as the unsheared prior, so the we must construct a 6-dimensional function
on a 5-dimensional space, and for each source galaxy integrate the
posterior over all 6 dimensions of \vecM.  Note that this BFD method
will be much faster than the BMF method even at equal dimensionality,
because BFD requires the evaluation of only one Gaussian
$\likeli(\vecM_i|\vecM)$ at each point of the integration, 
whereas BMF requires evaluation of the full likelihood of the pixel
data of a model.

Since galaxy-evolution theory will not be able in the foreseeable
future to provide an {\it a priori\/} distribution of galaxy
appearances, the prior will always have to be empirical.  
The prior can be established by obtaining high-$S/N$ data on a sample
of the sky.  Note that the \vecM\ values depend upon the choice of
weight function $W$ to suit the observational PSF, and
implicitly depend upon the filter passband used for the observations,
so the prior is best 
constructed from deep integrations using the same instrument as the
main survey.  The number of high-$S/N$ galaxies necessary to
adequately define the prior is an important issue for future study.

The prior can be generated by some kernel density estimation over the
empirical moments $\vecM_\mu$ where $\mu$ indexes the members of the
high-$S/N$ template galaxy set.  
In (\ref{posterior2}), however, we see that the prior is already being 
convolved with the likelihood function for a target galaxy.  Hence if the we are
using target galaxies of sufficiently low $S/N$ that $\likeli(\vecM_i
| \vecM)$ is broad enough to sample many elements of the template
set, it is adequate to express the prior as a sum
of delta functions for each template galaxy:
\begin{equation}
P(\vecM | \vecg) \propto \sum_\mu \delta^m\left[\vecM -
  \vecM_\mu(\vecg)\right],
\end{equation}
where we need
$\vecM_\mu(\vecg)$, the moments for the template galaxy if it were
sheared by $\vecg$. The posterior contribution for galaxy $i$ becomes
\begin{equation}
P(\vecD_i | \vecg) \propto \sum_\mu \likeli\left[\vecM_i | \vecM_\mu(\vecg)\right].
\end{equation}
We perform a Taylor expansion on this shear-conditioned probability to obtain our
required properties:
\begin{align}
P_i & =  \sum_\mu \likeli\left(\vecM_i | \vecM_\mu\right), \notag \\
\vecQ_i & =  \sum_\mu \left[ \frac{\partial}{\partial \vecM}
  \likeli\left(\vecM_i | \vecM \right)\right]_{\vecM_\mu} \cdot
  \bnabg \vecM_\mu, \\
\matR_i & =  \sum_\mu   \bnabg \vecM_\mu \cdot
 \left[ \frac{\partial^2}{\partial \vecM^2}
  \likeli\left(\vecM_i | \vecM \right)\right]_{\vecM_\mu} \cdot
  \bnabg \vecM_\mu \notag\\
 & +  \sum_\mu \left[ \frac{\partial}{\partial \vecM}
  \likeli\left(\vecM_i | \vecM \right)\right]_{\vecM_\mu} \cdot
  \bnabg \bnabg \vecM_\mu, \notag
\end{align}
where all moments and derivatives are taken at $\vecg=0$.
The determination of $\bnabg \vecM_\mu$ from the data for template
galaxy $\mu$ is straightforward in Fourier domain, and is given in
Appendix~B.

Now we adopt the multivariate Gaussian likelihood
(\ref{gausslikelihood}) for the moments.  We also can define
$\vecM_\mu(\vecx,\phi)$ to be the moments that we would assign to
template galaxy $\mu$ if it were translated to $\vecx$ (relative to
the target galaxy's coordinate origin) and rotated by $\phi$.  Because
the unlensed sky is isotropic, we can assume that the true prior
contains replicas of template galaxy $\mu$ at all \vecx\ and
$\phi$.\footnote{We can limit the potential positions 
  \vecx\ of the template galaxy centroid to be within the range of
  pixels assigned to our target galaxy $i$, essentially adopting a
  prior that we have indeed found a galaxy and its center is within
  our postage stamp.  This avoids the problem of divergent position
  marginalization noted in  \citet{lensfitcfh}.}  
The parity invariance of the unlensed sky also implies that we can
place a mirror image of each template galaxy in the prior as well.
Appendix~\ref{momentcalcs} shows
how to calculate moments for these transformed versions of a template
galaxy. 

For notational simplicity we will subsume the parity flip into the
integration over rotation $\phi$.   The Taylor-expanded posterior
derived from the template sample now becomes:
\begin{align}
P_i & =  \sum_\mu \int d\vecx\, d\phi\, \likeli_{i\mu}(\vecx,\phi) \notag \\ 
\label{final2}
\vecQ_i & = \sum_\mu \int d\vecx\, d\phi\, 
\likeli_{i\mu}(\vecx,\phi)
\left[ \vecM_i - \vecM_\mu(\vecx,\phi) \right] \cdot \matC_i^{-1} \cdot
\bnabg\vecM_\mu(\vecx,\phi) \\
\matR_i & =  \sum_\mu \int d\vecx\, d\phi\, 
\likeli_{i\mu}(\vecx,\phi) \left\{
\left[ \vecM_i - \vecM_\mu(\vecx,\phi) \right] \cdot \matC_i^{-1} \cdot
\bnabg\bnabg\vecM_\mu(\vecx,\phi) \right. \notag \\  
 &  \phantom{\sum_\mu \int d\vecx\, d\phi\, \likeli_{i\mu}(\vecx,\phi)}
\left. + \bnabg\vecM_\mu(\vecx,\phi) \cdot \matC_i^{-1}
\cdot \bnabg\vecM_\mu(\vecx,\phi) \right\} \notag \\
\likeli_{i\mu}(\vecx,\phi) & \equiv  \exp\left\{ -\frac{1}{2} 
\left[ \vecM_i - \vecM_\mu(\vecx,\phi) \right] \cdot \matC_i^{-1} \cdot
\left[ \vecM_i - \vecM_\mu(\vecx,\phi) \right] \right\}  \notag
\end{align}

The computational challenge of this Bayesian Fourier-domain shear inference
is the high multiplicity of
this calculation:  for every target galaxy $i$, we collect 6 sums over every
template galaxy $\mu$, with each term of each sum being an integral of
a Gaussian function of the three dimensions plus parity flip of
$(\vecx,\phi)$.  There are obvious efficiencies to be gained in this
calculation by pruning the template set to those with
significant contributions to the sums. Furthermore we recall that all
science will come from sums over a large number of target galaxies, so
we can subsample the template set when
computing the posterior for individual target galaxies, if we can
do so without inducing systematic biases on $(P_i, \vecQ_i, \matR_i)$.

A substantial speedup of the Bayesian shear calculation is enabled if
we can approximate $\vecM_\mu$ as linearly dependent on \vecx,
in which case two of the three dimensions of the integrals in
Equations~(\ref{final2}) reduce to linear algebra.
Appendix~\ref{momentcalcs} shows how to determine these
derivatives for the template galaxies.

We leave the testing of a practical implementation of
Equations~(\ref{final2}) and the investigation of the required size of
the template galaxy sample to further work.

\section{$P(\vecD_i | \vecg)$ via null tests}
\label{nulltests}
The BFD method improves on the BMF method by eliminating the 
approximation that target galaxies are described by a low-dimensional
model.  We paid a price, however, in losing the convenience of having
the action of shear be fully described by alteration of just the two
components of \vece.  This allowed us to derive the
lensed prior solely from derivatives with respect to $e$ of the
unlensed prior $P_0(e,\vect).$   In this Section we ask whether
null-testing methods can be used to make a model-free Bayesian shear
inference with the simplicity of a known shear transformation
$\vece\oplus\vecg.$  We conclude below that this is
  difficult.  Readers uninterested in the null-testing approach can
  safely skip this section.

Windowed centroiding procedures assign a center \vecx\ to a galaxy by
translating the galaxy until the windowed first moments are nulled.
The galaxy is assigned a centroid that is the inverse of the
translation needed to null the moments.  In 
the Fourier Domain Null Test (FDNT) method \citep{fdnt}, this is
extended by shearing as well as translating the galaxy (after
correcting for seeing) until we null the moments
$M_x, M_y, M_+,$ and
$M_\times$ in \eqq{moments}.  The galaxy is assigned a
shape $\vece_i$ that is the inverse of the shear that produces the
null.  The moment vector for galaxy $i$ is hence a function $\vecM_i(\vecE)$ of the
four-dimensional transformation $\vecE=(\vecg, \vecx)$ and we assign
the galaxy a shape and centroid $\vecE_i=(\vece_i, \vecx_i)$ such that
$\vecM_i(-\vecE_i)=0$. 
This approach
assures a well-determined transformation of the measured shape
$\vece_i$ under an applied shear \vecg, and
there is no need of a galaxy model.  \citet{fdnt}
demonstrates shear inferences errors of $<1$ part in $10^3$ on
low-noise data using FDNT.

The application of rigorous Bayesian formalism to FDNT is foiled,
however, because there is no straightforward means of propagating the
pixel noise model to a
likelihood $\likeli(\vecE_i | \vecE)$ of measuring a
null at $\vecE_i$ when the underlying galaxy has true null at
$\vecE$.  
It is possible, however, to approximate $\likeli(\vecE_i |
\vecE)$ in the case where the measured shape and centroid are
close to the true ones.  At sufficiently high $S/N$ of the target
galaxy, the likelihood will be confined to such regions and the
Bayesian formulation with this approximation will become accurate.

The measured moments $\vecM_i$ for galaxy $i$ when transformed by
$\vecE$ can be written as
\begin{equation}
\label{null1}
\vecM_i(\vecE) = \vecM_\mu(\vecE) + \delta \vecM_i(\vecE),
\end{equation}
where $\vecM_\mu$ is the true underlying moment and $\delta \vecM_i$ is
the variation induced by measurement noise.  Since $\vecM_i(\vecE)$ is
a linear function of the pixel data, the likelihood
$\likeli_i\left[\delta \vecM_i(\vecE)\right]$ is calculable and close to Gaussian.  

Our first approximation is to linearize $\vecM_\mu$ about the
$\vecE_\mu$ that nulls it:
\begin{align}
\vecM_\mu(\vecE) & =  \matD_\mu \cdot (\vecE-\vecE_\mu), \\
\matD_\mu & \equiv \left. \frac{\partial \vecM_\mu}{\partial
    \vecE}\right|_{\vecE_\mu}.
\end{align}
Second we assume that $\delta\vecM_i(\vecE)$ is invariant in the
neighborhood of $\vecE_i$ and that this measurement error has a 
multivariate Gaussian likelihood $\likeli_i(\delta\vecM)$ defined by a
covariance matrix $\matC_i$ and zero mean.
Then via \eqq{null1} the nulling transformation
$\vecE_i$ is
\begin{align}
 0 = \vecM_i(\vecE_i) & =  \matD_\mu \cdot \delta\vecE_{i\mu} + \delta
\vecM_i \\
\Rightarrow  \quad \likeli(\vecE_i | \vecE_\mu) & = 
\likeli_i(-\matD_\mu\cdot\delta\vecE_{i\mu}) \left| \matD_\mu \right|.
\end{align}

If we again construct the prior from a template set of galaxies
indexed by $\mu$, having ellipticities $\vece_\mu$, and uniformly
distributed in position $\vecx_\mu$ with respect to the 
target galaxy's position, then the contribution to the posterior from
target galaxy $i$ is the sum over template galaxies:
\begin{equation}
\label{null2}
P(\vecE_i | \vecg) \propto \sum_\mu \int d^2x_\mu\, 
\left| \matD_\mu \right| \exp\left[ -\frac{1}{2} \delta\vecE_{i\mu}^T
  \matD_\mu^T \matC_i^{-1} \matD_\mu \delta\vecE_{i\mu} \right].
\end{equation}
The integration over the two spatial dimensions $\vecx_\mu$ of the
Gaussian argument $\vecE_\mu=(\vece_\mu, \vecx_\mu)$ is analytic.  To
complete the implementation of the Bayesian framework we need to find
the Taylor expansion of \eqq{null2} with respect to \vecg.  The
right-hand side depends implicitly on the applied shear \vecg\ because
the template galaxy's shape $\vece_\mu$ is transformed by applied
shear and enters into $\delta\vecE_{i\mu}$.  To be thorough we should
also account for the variation of the derivative vector $\matD_\mu$
with applied shear.  Doing so is straightforward but not instructive
so we omit the algebra here.

We find therefore that there is a high-$S/N$ approximation to the
Bayesian shear estimator in the case of iterative null tests, but that
this requires knowing the prior distribution not only of the galaxies'
ellipticities $\vece$ but also of the derivatives $\matD$ of the null
tests with respect to shear and translation.  Hence even in the high-$S/N$
approximation, we have not found a way to produce a simpler model-free
derivatives of the prior than in BFD.
We will therefore choose to pursue BFD first, as it
demands no approximations beyond the weak-shear Taylor expansion of the
posterior and does not require searching for nulled moments.

\section{Comparison to other methods}
\label{compare}
\subsection{Model fitting}
The BMF shear measurement that we propose bears
very close resemblance to \lensfit.
\lensfit\ is described
as ``Bayesian galaxy shape measurement'' and differs 
from our BMF method which is derived as a Bayesian {\em shear}
determination.  The consequence is that BMF does not assign
shapes (ellipticities) to galaxies, instead computing the
likelihood-weighted derivatives of the prior for each galaxy as per
(\ref{int1}), and combining to yield shear as per (\ref{Cg}) and (\ref{barg}).

\lensfit\ already does the hard computational work of our BMF
algorithm, namely the integration of the data likelihood times the
prior over the parameters of the model space. Hence \lensfit\ serves as an
example of the feasibility of our BMF method.
At a minimum level of realism, the galaxy model must
include two ellipticity components \vece, two centroid components
\vecx, and \vect\ including a galaxy flux, size, and a measure of
concentration such as the Sersic index, for
7 parameters.  Isotropy reduces the unlensed prior to 6 dimensions.
The posterior requires integration over 7 dimensions.
\citet{lensfit} reduce the dimensionality of the likelihood integral
by analytic marginalization over flux (which requires a particular
choice of prior) and linearization of the model
dependence on \vecx.  \citet{lensfitcfh} use a bulge/disk ratio in place of a
Sersic index for concentration.  
Their work hence
demonstrates the feasibility of computing the Bayesian integrals over a 7-dimensional
model space.  The \lensfit\ implementations simplify the prior substantially by
separating the dependence on variables.
A numerical approach to constructing an accurate fully-coupled prior from
high-$S/N$ observations remains to be demonstrated.

There are reasons to suspect that this minimal model does not describe
the true galaxy population sufficiently well to achieve $|m|<10^{-3}$,
in particular because galaxies with radial gradients in ellipticity
induce biases in shear inferences when fit by models without gradients
\citep{fdnt}.  The computational difficulty will grow, with an
exponential increase in the number of likelihood evaluations, as additional
parameters are needed in the model and hence in the prior and the
integrations  This will lead us to favor the BFD method.  
Model-fitting will remain useful, however, in cases of
incomplete pixel information on the galaxy, {\it e.g.\/} from cosmic
rays, in which case the models can serve as a sparse dictionary for
the pixel data.

\subsection{Moment compression}
The Bayesian Fourier Domain (BFD) method is model-free, in common with
many moment-based schemes for shear inference, the best-known of which
is from \citet[KSB]{ksb}.  BFD shares with KSB and its brethren the
approach that the galaxy pixel data can be compressed to a few simple
moments which transmit most of the information on lensing shear of the
image.  The BFD method, however, differs fundamentally in explicit use
of a prior high-$S/N$ moment distribution, as opposed purely to
summations over properties of the normal-$S/N$ images.  There are
fundamental differences between BFD and KSB that give the former a
rigorous treatment of the effects of PSFs and noise:
\begin{itemize}
\item The BFD method accumulates moments in Fourier domain, admitting
  an exact correction for the joint effects of the PSF and shear on
  the images, avoiding the need for KSB's Gaussian-based
  approximations.
\item The BFD method does not require an iterative centroiding
  procedure, meaning the full likelihood of the observed moments remains known even
  at low $S/N$.  Positional uncertainties are treated by Bayesian integration
  over all possible positions of the galaxy.
\item The ``polarizability'' that KSB calculates for each galaxy is
  effectively replaced by integration of the target galaxy's moment
  likelihood over the shear derivatives of the prior.  The BFD
  approach is exact in the presence of noise whereas the
  single-galaxy polarizabilities are biased by noise.  A consequence
  of this is the absence of an ellipticity assignment to individual
  galaxies in BFD.
\end{itemize}

The non-Bayesian method bearing the closest resemblance to BFD is
described by \citet{Zhang11} and its predecessor papers. Like BFD,
Zhang's method is equivalent to accumulating moments in Fourier domain, and
emphasizes that a shear inferred from the ratio of two sums over the
galaxy population---an average ``signal'' divided by an average
``responsivity''---is less biased by noise than a shear inferred from a
sum of single-galaxy ratios {\it a la\/} KSB.  A critical difference
is that Zhang's method takes moments of the power spectrum $|\tilde
I^2(\veck)|$ rather than the Fourier amplitude $\tilde I(\veck)$.
This makes the method insensitive to the choice of centroid.  It also,
however, amplifies the noise relative to the signal and the rectified
noise must be very precisely removed from the power spectrum.
Neither Zhang nor KSB have an intrinsic means of appropriately
weighting the galaxies for the shear information that they contain,
which occurs naturally in the BFD method.  This means they require
some form of weighting or selection in order to keep the low-$S/N$
galaxies from ruining the $S/N$ of the shear inference.  Weighting and selection can
themselves induce biases in the inferred shear.
Finally, Zhang's method is formulated only for a Gaussian weight
function $W(|k|)$, which formally requires integration over regions of
$k$-space with infinite noise in any real image.  

High shear fidelity on simulated images has been achieved by
``stacking'' methods, which sum galaxy images before analysis
\citep{Kuijken,Lewis}. These work by essentially creating a single
high-$S/N$ image with an unlensed source that must approach
circularity. Noise-induced biases are therefore avoided and the number
of parameters needed to approximate the mean galaxy is greatly
reduced.  These stacking methods have practical drawbacks when the PSF
and/or the lensing shear vary across the image.  There is also a
deeper problem in the need to assign an origin to each galaxy
to build the stack.  At this stage the galaxies' individual low $S/N$
levels are still present and noise-induced biases, primarily a
suppression of the shear inferred from the stack, are important.  The BFD method will
avoid this problem.

\section{Conclusion}
\label{conclude}
We have shown how the exact Bayesian formulation of shear inference
from a collection of galaxy images can be turned into a practical
measurement technique in the limit of weak shear, by accumulating each
target galaxy's contribution to the first terms of the power-law expansion of the
posterior $\ln P(\vecg | \vecD)$.  We can expect this rigorously
derived treatment of noise to yield highly accurate shears even at low
$S/N$, and also make it possible to suppress selection biases in the
shear inference.  We confirm the ability of the algorithm to provide
part-per-thousand shear inference at low $S/N$, in the case of a
highly simplified model of galaxy measurement.

The Bayesian derivation leads to a shear estimator that departs in
significant ways from most previously proposed methods.  Ellipticities
(shapes) are not assigned to individual galaxies; biases are
avoided by combining a large ensemble of low-$S/N$ galaxies into a
single shear estimator.  Another important element is the
marginalization over galaxy position rather than selecting a
centroid.  These elements have been used seperately in previous
methods, but not together. 

There are two clear routes to coupling the Bayesian treatment of noise
with an exact treatment of the effects of the PSF and sampling on the
image: Bayesian Model Fitting (BMF) assumes the target galaxies follow
known parameterized forms, and the pixel data can be compared to
models that have the instrumental effects applied.  BMF needs to know
the distribution of the unlensed population over the model
parameters.  The \lensfit\ code has demonstrated the computational
feasibility of this approach for the minimal galaxy models.  Any
model-fitting approach is only accurate, however, insofar as the real
galaxies hew to the models.  Given the infinite variety of real
galaxies, there is an inherent approximation remaining in the BMF
method, which needs to be evaluated.  Model-fitting methods do however
have the
advantage of working even with incomplete or aliased pixel data for a galaxy.

The second route, Bayesian Fourier Domain (BFD) inference, compresses
both the target galaxies and the prior into a set of $k$-space
moments.  Exact corrections for PSF and sampling are possible and the
method is model-free if the images are fully sampled.  The likelihood
of the (compressed) data given the underlying moment vector is a
well-defined multivariate Gaussian in the common case of
background-limited observations.   We propose to apply the method to
the zeroth, first, and second moments of the galaxies.  Measuring
these 6 moments will be fast and foolproof---no iteration is required.  The prior
distribution of moments will be 5-dimensional because of isotropy.
We show how the prior can be constructed and integrated using
empirical moment measurements on high-$S/N$ images obtained from long
integrations on a small subset of the sky area in a given survey.
This integration is likely to be the most computationally intensive
step of an implementation of the BFD method.  The required size of
this template subsample is an important issue for future study.

Both the BMF and BFD methods require empirical inputs in the form of
distributions of real galaxies over some galaxy property vector
\vecG---galaxy ellipticities, Sersic indices, etc. in the BMF case,
and seeing-corrected galaxy moments in the BFD case.  Also in either
case one must know how this empirical distribution changes under
application of lensing shear (and magnification).  In the BMF case,
this typically is a simple shift of galaxy ellipticity parameters.  In
the BFD case, the derivatives of galaxy moments under shear are simply
higher-order moments derived in Appendix~\ref{momentcalcs}.  Hence BFD
is truly model-free in the sense that it is not necessary to
characterize the full surface-brightness distributions of real
galaxies---just the empirical distribution of a finite set of galaxy
moments.

Both the BMF and BFD methods easily treat the cases of multiple
exposures of a single galaxy, spatially varying shear and PSFs, and
are extensible to the inference of lensing magnification as well as
shear.  While we defer testing of implementations of these methods
to a future publication, the absence of approximations in the
derivation of these methods, particularly the BFD method, gives great
hope of achieving shear inference at better than part-per-thousand
accuracy, as is required for future ambitious weak lensing surveys.

\acknowledgements
Both authors were supported by Department of Energy grant DE-SC0007901.  GMB additionally acknowledges
support from National Science Foundation grant AST-0908027 and NASA
grant NNX11AI25G.  We are grateful to Sarah Bridle, 
Bhuvnesh Jain, Mike Jarvis, Marisa March, and Lance Miller for
conversations that have helped guide this work.
We thank the referee, whose comments on this submission led us to significantly
better presentation and understanding of the method described herein.

\newpage
\appendix
\section{Third-order posterior calculation}
\label{cubic}
Here we take the expansion for $\ln P(\vecg | \vecD)$ to third order
in $g$, and calculate the perturbation to the maximum and mean of the
posterior relative to the second order (Gaussian) result in
Section~\ref{weakapprox}.   We adopt a slightly different notation by
first expanding the transformed prior as
\begin{align}
\label{qrs}
P(\vece^o | \vecg) & =  P_0(\vece^o \oplus -\vecg)  
\left| \frac{d\vece^s}{d\vece^o} \right|_{-\vecg} \\
& =  P_0(e) + q(e) g \cos\Delta\phi \notag \\
 &\phantom{{}={}} + \frac{g^2}{2}\left[r_0(e) + r_2(e) \cos 2\Delta\phi\right] \notag \\
 &\phantom{{}={}} + \frac{g^3}{6}\left[ s_1(e) \cos\Delta\phi + s_3(e) \cos
   3\Delta\phi \right].
\end{align}
Both the unlensed prior and the shear transformation formula must be
invariant under rotation of coordinates, so
the sheared prior must be a function only of the ellipticity amplitude
$e = | \vece^o |$, the shear amplitude $g=|\vecg|$, and the angle
$\Delta\phi$ between them.  Sine terms are absent because the shear
transformation law should be invariant under parity flip.

The functions $q, r_0, r_2, s_1,$ and $s_3$ of galaxy ellipticity $e$ can be expressed as
algebraic combinations of derivatives of $P_0$ and of the shear
addition law.  These expressions are complex and not enlightening, and in practice
a numerical estimate is likely to be faster than the algebraic
calculation anyway, so we omit the algebraic forms.  Throughout this
Appendix we also omit the additional galaxy parameters \vect\ which
will also be arguments of $P_0$ and the five derivative functions.

Adopting the complex notation $g=g_x + i g_y$, and designating $\phi$
as the azimuthal angle of \vece, the posterior likelihood of shear for
galaxy $j$ becomes
\begin{align}
\label{cubicPi}
P(\vecg | \vecD_i) & \propto  P_i + {\rm Re} \left[ Q_j^\ast g
 + \frac{1}{2}\left(R_{0j}^\ast gg^\ast + R_{2j}^\ast g^2\right) + \frac{1}{6}\left(S_{1j}^\ast g^2g^\ast +
 S_{3j}^\ast g^3\right) \right], \\
\left\{ \begin{array}{c}
P_j \\
Q_j \\
R_{0j} \\
R_{2j}\\
S_{1j} \\
S_{3j} \\
\end{array} \right\} & \equiv 
\int e\,de \, d\phi \, P(\vecD_j | \vece)
\left\{\begin{array}{c}
P_0( e) \\
q(e) \exp(i\phi) \\
r_0(e)  \\
r_2(e) \exp(2i\phi) \\
s_1(e) \exp(i\phi) \\
s_3(e) \exp(3i\phi) \\
\end{array} \right\} \notag
\end{align}

After taking the logarithm of the posterior in (\ref{cubicPi}) to
third order in $g$, we sum over galaxies to get the (log of) the total
shear posterior probability:
\begin{align}
\label{cubicposterior}
-\ln P(\vecg | \vecD) & =  -\sum \ln P_j + {\rm Re} \left[ Q^\ast g
 + \frac{1}{2}\left(R_0^\ast gg^\ast + R_2^\ast g^2\right)
 + \frac{1}{6}\left(S_1^\ast g^2g^\ast +
 S_3^\ast g^3 \right) \right], \\
-Q & =  \sum \frac{Q_j}{P_j}, \label{qsum} \\
-R_0 & =  \sum \left( \frac{R_{0j}}{P_j} 
  - \frac{1}{2}\frac{Q_j Q_j^\ast}{P_j^2}\right), \\
-R_2 & =  \sum \left( \frac{R_{2j}}{P_j} 
  - \frac{1}{2}\frac{Q_j^2}{P_j^2}\right), \\
-S_1 & =  \sum \left( \frac{S_{1j}}{P_j} 
  - 3\frac{R_{0j} Q_j}{P_j^2}
  - \frac{3}{2}\frac{R_{2j}  Q_j^\ast}{P_j^2}
  + \frac{3}{2}\frac{Q_j^2 Q_j^\ast}{P_j^3}\right), \\
-S_3 & =  \sum \left( \frac{S_{3j}}{P_j} 
  - \frac{3}{2}\frac{R_{2j}  Q_j}{P_j^2}
  + \frac{1}{2}\frac{Q_j^3}{P_j^3}\right). \label{s3sum}
\end{align}
All sums are over the source galaxy index $j$.  Note that in the
absence of applied shear, and with a circularly symmetric PSF, there
is no preferred direction in the \vece\ plane, and the expectation
value of $Q, R_2, S_1,$ and $S_3$ are all zero.  Only $R_0$,
therefore, remains finite in this limit.  The other sums will be at
most $O(g)$, although we will retain all terms in case the PSF can
break the symmetry of the unsheared observations.

The posterior (\ref{cubicposterior}) is simpler if we rotate to the
frame where $R_2$ is real and the covariance matrix is diagonal.  If
the original $R_2$ has phase $2\phi$, then we set
\begin{align}
\label{cubicrotate}
\tilde g & =  g e^{-i\phi}  \equiv  x + iy \\
\tilde Q & =  Q e^{-i\phi} \equiv Q_x + i Q_y \notag \\
\tilde S_1 & =  S_1 e^{-i\phi} \equiv S_{1x}+iS_{1y} \notag \\
\tilde S_3 & =  S_3 e^{-3i\phi} \equiv S_{3x}+iS_{3y} \notag \\
\sigma^2_x & =  \left( R_0 + |R_2|\right)^{-1}  \notag \\
\sigma^2_y & =  \left( R_0 - |R_2|\right)^{-1}. \notag
\end{align}
As an aside, we note that if the measurement process (including PSF)
has no preferred direction, then the applied shear \vecg\ sets the
only axis of the system (which is $\phi$), and a consequence will be
that $\langle \tilde Q \rangle = \langle \tilde S_1 \rangle = \langle
\tilde S_3\rangle=0$.

At quadratic order, (\ref{cubicposterior}) implies a Gaussian
posterior distribution in \vecg\ with both maximum and mean at
\begin{equation}
\label{quadraticsoln}
(x_0,y_0) = (-\sigma^2_x Q_x , -\sigma^2_y Q_y) = \left (
\frac{-Q_x}{R_0+|R_2|}, \frac{-Q_y}{R_0-|R_2|} \right)
\end{equation}
and independent errors on the components $x$ and $y$ of the shear in
this rotated frame, equivalent to equations~(\ref{barg}) and
(\ref{Cg}).  We treat the cubic terms in (\ref{cubicposterior}) as
perturbations to the Gaussian by assuming the $|S_{1,3} g^3| \ll 1$
at any $g$ for which the Gaussian likelihood is significant.  This
approximation can be restated as
\begin{align}
| S Q^3 R_0^{-3} | & \ll 1, \\
| S R_0^{-3/2} | & \ll  1. 
\end{align}
Under these conditions the posterior likelihood in the rotated \vecg\
frame can be expanded to first order in $S_{1,3}$ giving
\begin{equation}
P(\vecg | \vecD) \propto 
\exp\left[ \frac{-(x-x_0)^2}{2\sigma^2_x}\right]
\exp\left[ \frac{-(y-y_0)^2}{2\sigma^2_y}\right]
\left\{ 1 - \frac{1}{6} {\rm Re} \left[ \tilde S_1^\ast (x+iy)
 + \tilde S_3^\ast (x+iy)^3\right]\right).
\end{equation}
The cubic term shifts the peak of the Gaussian posterior slightly, and 
adds some skew which shifts the expectation value further.  The expectation
value of shear under this posterior can be integrated to give
\begin{align}
\label{expectation}
\langle x \rangle = {} & x_0 
- \frac{ S_{1x} + S_{3x}}{R_0+|R_2|}
\left( \frac{ \sigma^2_x + x_0^2}{2}\right)
- \frac{ S_{1x} - S_{3x}}{R_0+|R_2|}
\left(\frac{ \sigma^2_y + y_0^2}{6}\right)  \\
 & \phantom{x_0}
- \frac{ S_{1y} - 3S_{3y}}{R_0+|R_2|}
\left(\frac{ x_0 y_0 }{3}\right) \notag \\
\langle y \rangle  = {} & y_0 
- \frac{ S_{1x} - 3S_{3x}}{R_0 -|R_2|}
\left(\frac{ x_0 y_0 }{3}\right) \notag \\
 & \phantom{y_0}
- \frac{ S_{1y} + S_{3y}}{R_0-|R_2|}
\left(\frac{ \sigma^2_x + x_0^2}{2}\right)
- \frac{ S_{1y} - S_{3y}}{R_0-|R_2|}
\left(\frac{ \sigma^2_y + y_0^2}{6}\right). \notag
\end{align}
In each case, the $\sigma^2$ terms in parentheses arise from the
skewness imposed on $P(\vecg | \vecD)$.  Omitting them gives the
location of the maximum of the posterior.  

To obtain some understanding of this result, we take the case when
the mean shear is
determined to high accuracy, so the skewness becomes unimportant.  Since
we expect $y_0, S_{1y},$ and $ S_{3y}$ to be smaller than their $x$
counterparts, and also $|R_2| \ll R_0$, the perturbation from cubic
terms is expected to be dominated by a shift of the Gaussian
likelihood to
\begin{align}
\label{simple}
\langle x \rangle & =  \frac{-Q_x}{R_0+|R_2|}
- \frac{ (S_{1x} + S_{3x})Q_x^2}{2(R_0+|R_2|)^3} \\
\langle y \rangle & =  \frac{-Q_y}{R_0-|R_2|}. \notag
\end{align}
The main effect of the cubic term is therefore a slight shift of the shear
posterior toward or away from the origin. We have not verified whether
this abbreviated form is sufficiently accurate
in cases with anisotropic PSFs, so the full form (\ref{expectation})
should be used.

To summarize, the procedure for estimating the shear using terms to
3rd order in $g$ is:
\begin{enumerate}
\item From the unlensed ellipticity distribution $P_0$ and the shear
  transformation law, calculate the functions $q, r_0, r_2, s_1,$ and
  $s_3$ of $e$ defined in equation~(\ref{qrs}).
\item For each source galaxy $j$, integrate over \vece\ as per
  equations~(\ref{cubicPi}) to obtain $P_j, Q_j, R_{0j}, R_{2j}, S_{1j},$
  and $S_{3j}$.
\item Sum over source galaxies as per equations~(\ref{qsum})--(\ref{s3sum}) to get $Q,
  R_0, R_2, S_1$, and $S_3$.
\item Rotate by the phase of $R_2$ to diagonalize the covariance
  matrix of the shear posterior and obtain transformed.  Phases of the
  $QRS$ components are changed and the shear component variances
  $\sigma^2_x$ and $\sigma^2_y$ are obtained as per
  equations~(\ref{cubicrotate}).
\item The maximum and mean of the quadratic solution in the rotated
  system are given by equations~(\ref{quadraticsoln}).
\item The perturbation to the expectation value of the posterior due
  to cubic terms is in equations~(\ref{expectation}).
\item Rotate back to original coordinate system.
\end{enumerate}

\section{Derivatives and transformations of the Fourier-domain moments}
\label{momentcalcs}
To implement the Bayesian Fourier-domain shear technique of
Section~\ref{method2}, we need to calculate the first two
derivatives of the moment vector in \eqq{moments} with
respect to shear for each template galaxy.  It is also helpful to know
how these moments change under rotation, translation, and
parity transformations, since the symmetries of the unlensed sky
suggest that any template galaxy can be replicated with these
transformations.

We start by writing the moments generally as
\begin{equation}
M_\alpha = \int d^2k\, \tilde I(\veck) W(|\veck^2|) F_\alpha(\veck).
\end{equation}
$\tilde I(\veck)$ is the Fourier transform of the pre-seeing galaxy
image, and is related to the sky-plane surface brightness $I(\vecx)$ by the usual
\begin{equation}
\tilde I(\veck) = \int d^2x\, I(\vecx) \exp(i\veck\cdot\vecx).
\end{equation}
Consider a new galaxy image which is an affine transformation of the
original image, specified by a linear amplification
$\matA$ followed by translation by $\vecx_0$:
\begin{equation}
I^\prime(\vecx) = I\left(\matA^{-1}\vecx-\vecx_0\right)
\end{equation}
Standard Fourier manipulations give
\begin{align}
\tilde I^\prime(\veck) & = |\matA| e^{i\veck^\prime\cdot\vecx_0} \tilde
I\left(\veck^\prime\right) \\
\veck^\prime &\equiv \matA^T \veck.
\end{align}
The transformed moments are
\begin{align}
M^\prime_\alpha & = |\matA| \int d^2k\, \tilde I(\veck^\prime)
W(|\veck^2|) F_\alpha(\veck)
e^{i\veck^\prime\cdot\vecx_0} \\
 &= \int d^2k \tilde I(\veck)
W\left[\left| \left(\matA^T\right)^{-1}\veck\right|^2\right]
F_\alpha\left[ \left(\matA^T\right)^{-1}\veck\right]
e^{i\veck\cdot\vecx_0}.
\label{momenttransform}
\end{align}
\subsection{Shear derivatives}
We define a two-component shear $\vecg=(g_1,g_2)$ of a galaxy image
with the flux-conserving transformation 
\begin{equation}
\matA^{-1} = 
\frac{1}{\sqrt{1-g^2}} \left( \begin{array}{cc}
1 + g_1 & g_2 \\
g_2 & 1 - g_1
\end{array}
\right).
\end{equation}
It is convenient to adopt a complex notation at this point:
\begin{align}
k & \equiv k_x + ik_y  &\partial & \equiv \frac{\partial}{\partial g_1} - i \frac{\partial}{\partial
  g_2}  \notag \\
g & \equiv g_1 + i g_2 &
\bar \partial &\equiv \frac{\partial}{\partial g_1} + i \frac{\partial}{\partial
  g_2} 
\end{align}
With this notation the action of shear $\veck\rightarrow
\left(\matA^T\right)^{-1}\veck$ becomes
\begin{equation}
k \quad \rightarrow \quad k^\prime = \left( 1- g\bar g\right)^{-1/2} \left( k + g \bar
  k\right).
\label{shearcomplex}
\end{equation}
\eqq{momenttransform} can now be restated in the complex
notation for the case of a shear transformation:
\begin{equation}
\label{momentcomplex}
M^\prime_\alpha = \int d^2k\, \tilde I(k) W(k^\prime \bar k^\prime) F_\alpha(k^\prime)
\end{equation}
We are interested in the 2 scalar and 2 complex moments defined as
\begin{align}
M_0 & = M_I & F_0 & =1 \notag \\
M_1 & = M_x + i M_y & F_1 & = ik \notag \\
M_2 & = M_+ + i M_\times & F_2 & = k^2 \notag \\
 & M_r & F_r & = k\bar k.
\end{align}
The shear derivative operators can be rewritten as
\begin{align}
\bnab_g & = \vecv \partial + \bar\vecv \bar \partial
& \vecv & \equiv \left(\begin{array}{c}
1 \\ i
\end{array}\right) \\
\bnab_g\bnab_g & = \vecv\vecv^T \partial^2 + \bar\vecv
\bar\vecv^T\bar \partial^2
+ 2 \matI_2 \partial \bar\partial
& \matI_2 & \equiv \left(\begin{array}{cc}
1 & 0 \\ 0 & 1
\end{array}\right)
\end{align}
Now the derivatives of the moments with respect to
shear are obtained by applying these operators to the moment
definition (\ref{momentcomplex}) after substituting in the shear wavevector
transformation (\ref{shearcomplex}).  For each moment, the derivatives
can be expressed as
\begin{align}
\bnab_g M_\alpha & = \int d^2k\, \tilde I(k) \left[ W(k\bar k)
  A_\alpha(k)  + W^\prime(k\bar k) B_\alpha(k) \right] \\
\bnab_g\bnab_g M_\alpha & = \int d^2k\, \tilde I(k) \left[ W(k\bar k)
  C_\alpha(k)  + W^\prime(k\bar k) D_\alpha(k) +
W^{\prime\prime}(k\bar k) E_\alpha(k) \right].
\label{mDerivs}
\end{align}
Table~\ref{mTable} summarizes the results of propagating the shear
derivatives into our weight and moment functions.  All of the
moments and their derivatives are simple weighted polynomial moments
of the galaxy Fourier transform.

\begin{deluxetable}{ccccc}
\tablewidth{0pt}
\tablecaption{
Functional forms of the integrands for moments and their derivatives,
as defined by Equations~(\ref{mDerivs}).  The derivatives of the
moments under translate in $x$ and $y$ directions are found by adding
factors of $i(k+\bar k)/2$ and $(k-\bar k)/2$ to the entries, respectively.
\label{mTable}
}
\tablehead{
\colhead{Moment} & \colhead{$M_0$} 
& \colhead{$M_1$} 
& \colhead{$M_2$} 
& \colhead{$M_r$} }
\startdata
$F_\alpha=$ & 
$1$ & $ik$ & $k^2$ & $k\bar k$ \\ \tableline

$A_\alpha = \vecv \times $ 
& 0 & $i \bar k$ & $2k\bar k$ & $\bar k^2$ \\
\phantom{$A_\alpha$}$+\bar\vecv \times$ 
& 0 & 0 & 0  & $k^2$ \\ \tableline

$B_\alpha = \vecv \times $
& $\bar k^2$ & $ik\bar k^2$ & $k^2\bar k^2$ & $k \bar k^3$ \\
\phantom{$B_\alpha$}$+\bar\vecv \times$ 
& $k^2$ & $ik^3$ & $k^4$ & $k^3 \bar k$ \\ \tableline

$C_\alpha = \matI_2 \times $ 
& 0 & $ik$ & $2k^2$ & $4k\bar k$ \\
\phantom{$C_\alpha$} $+\vecv \vecv^T \times $ 
& 0 & 0 & $2\bar k^2$ & 0 \\ \tableline

$D_\alpha = \matI_2 \times $ 
& $4k\bar k$ & $6ik^2\bar k$ & $ 8k^3 \bar k$ & $8k^2\bar k^2$ \\
\phantom{$D_\alpha$}$+ \vecv\vecv^T \times $ 
& 0 & $2i\bar k^3$ & $ 4k \bar k^3$ & $2\bar k^4$ \\
\phantom{$D_\alpha$}$+ \bar\vecv \bar\vecv^T \times $ 
& 0 & 0 & 0 & $2k^4$ \\ \tableline

$E_\alpha = \matI_2 \times $ 
& $2k^2\bar k^2$ & $2ik^3 \bar k^2$ & $2k^4 \bar k^2$ & $2k^3\bar k^3$ \\
\phantom{$E_\alpha$}$+ \vecv\vecv^T \times $ 
& $\bar k^4$ & $ik\bar k^4$ & $k^2\bar k^4$ & $k \bar k^5$ \\
\phantom{$E_\alpha$}$+ \bar\vecv \bar\vecv^T \times $ 
& $k^4$ & $ik^5$ & $k^6$ & $k^5 \bar k$
\enddata
\end{deluxetable}

\subsection{Rotation transformations}
In our adopted complex notation, the effect of rotating the galaxy by
angle $\phi$ is to send $k\rightarrow e^{i\phi}k$ in the argument of
$F_\alpha$ in \eqq{momenttransform}.  The monopole moments
$M_0$ and $M_r$ are unchanged, while the dipole and quadrupole moments
$M_1$ and $M_2$ acquire factors $e^{i\phi}$ and $e^{2i\phi}$,
respectively.  The rotational behavior of all the shear derivatives of the
moments can also be easily assessed by applying the phase factors to
all powers of $k$ and $\bar k$ in the elements of Table~\ref{mTable}.

\subsection{Parity transformations}
A parity flip sends $k \leftrightarrow \bar k$ for all of the
integrands of the moments and their derivatives in
Table~\ref{mTable}.  The moments $M_1$ and $M_2$ are conjugated,
meaning that $M_y$ and $M_\times$ change sign, while the other moments
are unchanged.  

\subsection{Derivatives with translation}
A translation of the galaxy by $\vecx_0$ adds a factor
$e^{i\veck\cdot\vecx_0}$ to $\tilde I(\veck)$ in the integrand of all
the moments (and their derivatives).  In general the integrations of
all the moments and their shear derivatives would need to be repeated.

We may, however, elect to approximate the effect of translation on the
moments and their shear derivatives by linearizing about $\vecx_0=0$.
In this case we are interested in $\frac{\partial M_\alpha}{\partial
  x_0}$, etc.  The derivative of any moment or its derivatives with
respect to $x_0$ can be obtained by adding a factor of $ik_x =
i(k+\bar k)/2$ to all the functions in Table~\ref{mTable}.  Similarly
the $y_0$ derivative adds a factor $ik_y = (k - \bar k)/2$ to the
integrands. 

If we adopt the linearization of the moments with translation, then
for each template galaxy we need a substantial number of real-valued
quantities: (6 moments) $\times$ (1 moment $+$ 5 derivatives for
\vecQ\ and \matR) $\times$ (1 value $+$ 2 translation derivatives) for
a total of 108 numbers per template.  The actual number of integrals
we need to perform is much less than this since many elements are
repeated in Table~\ref{mTable}.  Recall too that these template moments that
define the prior only need to be evaluated once for the experiment,
and the integrations are expressed as standard linear algebra
routines.  Evaluation of these quantities will not be a significant
computational burden.

\end{document}